\documentclass[journal]{IEEEtran}

\usepackage{xcolor}
\usepackage{acronym}
\usepackage{graphicx}
\usepackage{amsmath}
\usepackage{tabularx,booktabs}
\usepackage{multicol}
\usepackage{makecell}
\usepackage{amsfonts}
\usepackage{bm}
\usepackage{tablefootnote}
\usepackage{subcaption}
\usepackage{url}
\usepackage{caption}
\usepackage{amssymb}
\usepackage{amsthm}
\usepackage{chngcntr}
\usepackage{parskip}
\newcolumntype{Y}{>{\centering\arraybackslash}X}
\newcolumntype{?}{!{\vrule width 1.5pt}}
\usepackage[numbers,sort&compress]{natbib}
\acrodef{PMF}[PMF]{probability mass function}
\acrodef{MAC}[MAC]{medium access control}
\acrodef{IRSA}[IRSA]{irregular repetition slotted ALOHA}
\acrodef{i.i.d.}[i.i.d.]{independent and identically distributed}
\acrodef{DL}[DL]{deep learning}
\acrodef{DNN}[DNN]{deep neural network}
\acrodef{BN}[BN]{Bayesian network}
\acrodef{MC}[MC]{Markov chain}
\acrodef{MAB}[MAB]{multi-armed bandit}
\acrodef{IoT}[IoT]{internet of things}
\acrodef{URLLC}[URLLC]{ultra-reliable low-latency communication}
\acrodef{MARL}[MARL]{multi-agent reinforcement learning}
\acrodef{RL}[RL]{reinforcement learning}
\acrodef{mMTC}[mMTC]{massive machine-type communication}
\acrodef{ACK}[ACK]{acknowledgment}
\acrodef{CTDE}[CTDE]{centralized-training / decentralized-execution}
\acrodef{GNN}[GNN]{graph neural network}
\acrodef{GA}[GA]{genetic algorithm}
\acrodef{IDQN}[IDQN]{independent deep Q-Network }

\theoremstyle{definition}

\usepackage{algorithm}
\usepackage{algpseudocode}
\algnewcommand{\LeftComment}[1]{\Statex \(\triangleright\) \textit{#1}}

\usepackage{soul,color}

\newtheorem{theorem}{Theorem}

\newtheorem{definition}{\hspace{0pt}\bf Definition}

\usepackage{ragged2e}
\usepackage{blindtext}

\newcommand{\blue}[1] {{\color{black}{#1}}}

\newcommand{\deniz}[1] {{\color{black}{#1}}}

\DeclareMathOperator*{\argmax}{argmax}
\DeclareMathOperator*{\maxb}{max}
\DeclareMathOperator*{\minb}{min}

\usepackage{titlesec}
\titlespacing\section{0pt}{5pt plus 4pt minus 2pt}{2pt plus 2pt minus 2pt}
\titlespacing\subsection{0pt}{5pt plus 4pt minus 2pt}{2pt plus 2pt minus 2pt}

\begin{document}

\title{Learning a Decentralized Medium Access Control Protocol for Shared Message Transmission} 

\author{Lorenzo~Mario~Amorosa,~\IEEEmembership{Member,~IEEE},~Zhan~Gao,~\IEEEmembership{Graduate Student Member,~IEEE},\\
Roberto~Verdone,~\IEEEmembership{Senior Member,~IEEE},~Petar~Popovski,~\IEEEmembership{Fellow,~IEEE},~and~Deniz~G\"und\"uz,~\IEEEmembership{Fellow,~IEEE}% <-this % stops a space

\thanks{
\indent This work was supported, in part, by the Velux Foundation, Denmark, through the Villum Investigator Grant WATER, nr. 37793. We acknowledge funding from the UKRI for the projects AI-R (ERC Consolidator Grant, EP/X030806/1) and SNS JU project 6G-GOALS under the EU’s Horizon program (Grant Agreement No. 101139232).\\
\indent L.M. Amorosa and R. Verdone are with the Department of Electrical, Electronic and Information Engineering (DEI), ``Guglielmo Marconi", University of Bologna 40136, Italy \& WiLab - National Wireless Communication Laboratory (CNIT), Bologna 40136, Italy. E-mail: \{lorenzomario.amorosa, roberto.verdone\}@unibo.it\\
\indent Z. Gao is with the Department of Electrical and Electronic Engineering, Imperial College London, London SW7 2AZ, U.K. \& the Department of Computer Science and Technology, University of Cambridge, Cambridge CB3 0FD, U.K. \emph{(Corresponding author.)} E-mail: zg292@cam.ac.uk\\
\indent P. Popovski is with the Department of Electronic Systems, Aalborg University, 9220 Aalborg, Denmark. E-mail: petarp@es.aau.dk\\
\indent D. G\"und\"uz is with the Department of Electrical and Electronic Engineering, Imperial College London, London SW7 2AZ, U.K. E-mail: d.gunduz@imperial.ac.uk\\
}
}

% header
\markboth{}{}

% make the title area
\maketitle

\begin{abstract}
In large-scale Internet of things networks, efficient medium access control (MAC) is critical due to the growing number of devices competing for limited communication resources. In this work, we consider a new challenge in which a set of nodes must transmit a set of shared messages to a central controller, without inter-node communication or retransmissions. \deniz{To our knowledge, this is the first work to formalize and analyze the multi-shared-message decentralized random access problem}. Messages are distributed among random subsets of nodes, which must implicitly coordinate their transmissions over shared communication opportunities. The objective is to guarantee the delivery of all shared messages, regardless of which nodes transmit them. We first prove the \deniz{existence of an optimal} deterministic strategy, and characterize the success rate degradation of a deterministic strategy under dynamic message-transmission patterns. To solve this problem, we propose a decentralized learning-based framework that enables nodes to autonomously synthesize deterministic transmission strategies aiming to maximize message delivery success, together with an online adaptation mechanism that maintains stable performance in dynamic scenarios. Extensive simulations validate the framework's effectiveness, scalability, and adaptability, demonstrating its robustness to varying network sizes and fast adaptation to dynamic changes in transmission patterns, outperforming \deniz{state-of-the-art} approaches.
\end{abstract}

\begin{IEEEkeywords}
Decentralized Learning, Distributed Coordination, Internet of Things, Medium Access Control
\end{IEEEkeywords}

\IEEEpeerreviewmaketitle

\section{Introduction}
\label{sec:introduction}

\IEEEPARstart{W}{ith} exponentially growing \deniz{number of} devices in modern wireless \Ac{IoT} networks, \ac{MAC} to coordinate multiple nodes contending for shared communication opportunities (in time, frequency, etc.) is \deniz{becoming} increasingly 
challenging, especially under tight resources and large node counts \cite{macSurvey, cheng2018industrial, macSurvey2}. 
Distributed random access strategies like slotted ALOHA are preferred for minimizing signaling overhead and avoiding centralized scheduling \cite{jian2016random}. These advantages scale with network size, although collisions, i.e., the risk that two or more nodes attempt to access the same transmission resource, remain a significant challenge. 
This challenge becomes more complex 
in large-scale \ac{IoT} deployments for tasks such as critical infrastructure monitoring. When multiple nodes report the same event, e.g., multiple sensors alarm on an industrial anomaly \cite{pdm2, pdm1, longhi}, handling resource contention in a collision-free manner becomes critical.

In this work, we consider a complex problem, in which a set of nodes must deliver a set of \textit{shared} messages to a central controller, without inter-node communication or retransmissions. 
Unlike conventional random access with independent data, shared messages introduces unique complexity. Specifically, these messages are distributed among random subsets of nodes, indicating that multiple nodes may simultaneously attempt to transmit the same message over shared communication opportunities. 
We refer to the nodes that have to send one or more messages in the current time slot as the \textit{active nodes}, and all the nodes that have to send the same message belong to the same \textit{active set}. 
Complexity arises because nodes need to  
coordinate who should transmit a shared message and over which transmission opportunity 
without retransmissions or central coordination, and the network must rely on implicit mechanisms to ensure each shared message is delivered successfully. 
\blue{This is further complicated by the fact that these active sets are completely unknown to the individual nodes.} 
In this case, a practical mitigation is permitting \emph{message repetition} across communication opportunities (e.g., IRSA-type schemes \cite{irsa}) to increase the chance that at least one copy of each shared message finds a collision-free opportunity. 

\begin{figure}[t]
    \centering
    \includegraphics[trim= {0 0 0 0}, clip, width=0.91\columnwidth]{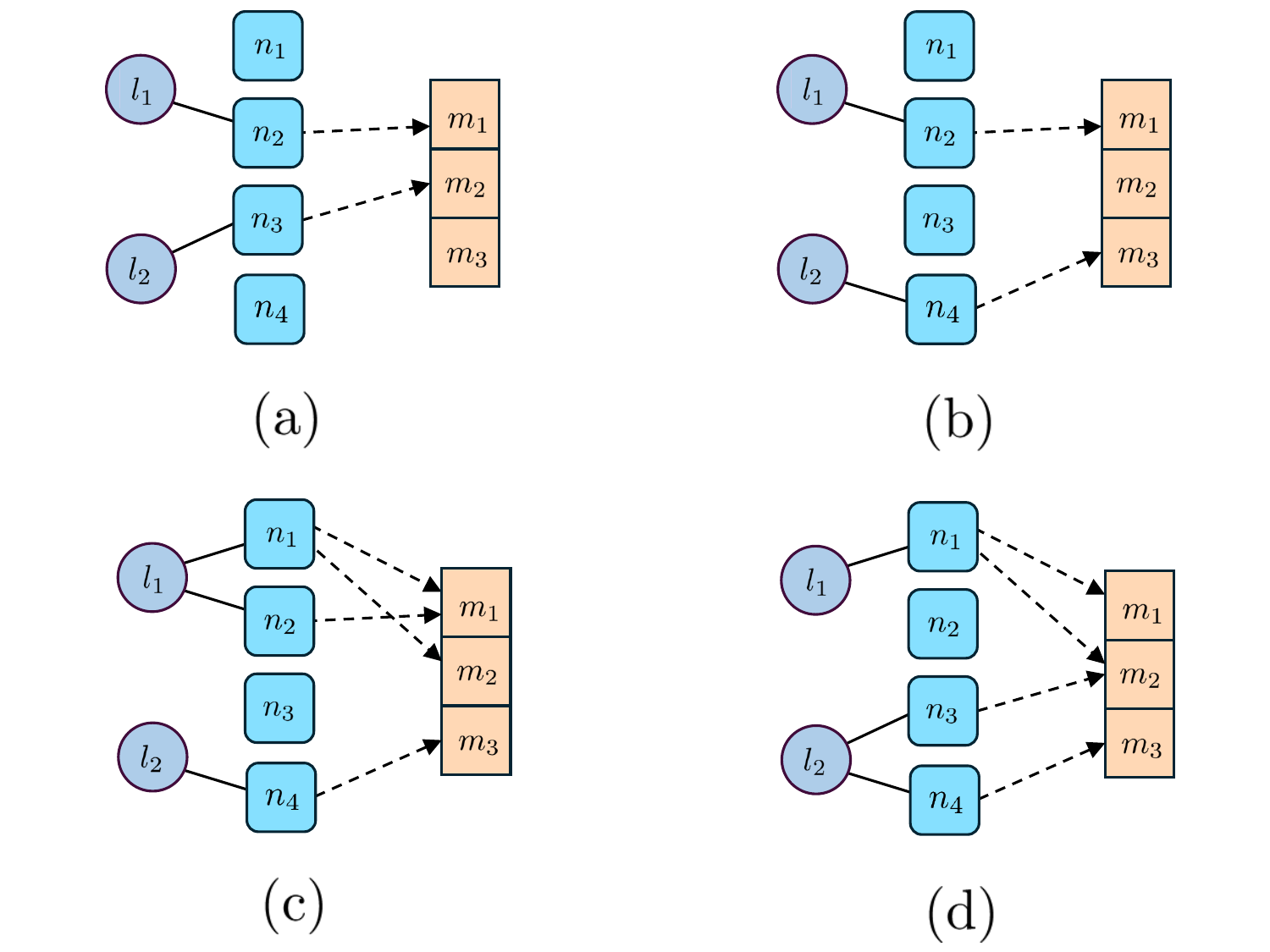}
    \caption{Example scenario where message repetition is crucial. Panels (a)-(d) demonstrate how a strategy with message repetition handles four possible active sets of shared messages with limited communication opportunities, 
    ensuring that for every active set at least one copy of each shared message can be delivered without collision.}
    \label{fig:repetition_example}
\end{figure}

Fig.~\ref{fig:repetition_example} shows an example where message repetition is  
crucial.  
Consider four nodes and two shared messages, $l_1$ (possibly available to nodes $n_1, n_2$) and $l_2$ (possibly available to nodes $n_3, n_4$). There are three shared transmission opportunities. 
We further assume that, at each time slot $t$, the set of active nodes can only be one of $\{n_2, n_3\},\ \{n_2, n_4\},\ \{n_1, n_2, n_4\},\ \text{or}\ \{n_1, n_3, n_4\}$.
We show by contradiction that, under such an %these 
activity pattern, any strategy that restricts each node to transmit its message on a single opportunity will fail, and message repetition is required to guarantee the successful delivery of all shared messages. 
Assume there exists a successful strategy in which every node transmits at most once at each $t$ (i.e., each node picks exactly one opportunity). We consider different possible active sets step by step: 
(i) When the active set is $\{n_1, n_2, n_4\}$, nodes $n_1$ and $n_2$ must ensure $l_1$ is received. 
Because there is no retransmission or inter-node coordination, and neither one node knows whether the other holds $l_1$, $n_1$ and $n_2$ must both attempt to transmit, yet they cannot do so on the same opportunity (as that would cause a self-collision for $l_1$). 
Similarly, $n_4$ (which carries $l_2$) must avoid colliding with the transmissions of $n_1$ and $n_2$ to deliver $l_2$ successfully. With only three opportunities available, this forces $n_1$, $n_2$, and $n_4$ to occupy three distinct opportunities; (ii) Next 
consider the active set $\{n_1, n_3, n_4\}$.  
It follows the same \deniz{reasoning} as in (i) that $n_3$ and $n_4$ must attempt to transmit, but cannot use the same opportunity as that would result in a self-collision for $l_2$. 
Moreover, $n_1$ (carrying $l_1$) must use an opportunity different from those chosen by $n_3$ and $n_4$ to prevent an inter-message collision. Because $n_1$ and $n_4$ already had fixed (distinct) opportunities from step (i), these constraints force $n_3$ to choose the same opportunity that $n_2$ used in step (i); (iii) Now consider the active set $\{n_2, n_3\}$. From the assignments deduced above, $n_2$ and $n_3$ have been driven to the same opportunity, so they would collide when both are active. That collision would cause both messages to fail delivery at that time slot $t$, contradicting the assumption that the single-transmission strategy always succeeds. In contrast, as shown in Fig. \ref{fig:repetition_example}, a strategy that allows controlled message repetition, i.e., assigning $n_1$ to transmit on both opportunity 1 and opportunity 2, $n_2$ on opportunity 1, $n_3$ on opportunity 2, and $n_4$ on opportunity 3, ensures that for every possible active set at least one copy of each shared message finds a collision-free opportunity.

Recognizing the necessity of message repetition, our goal is to maximize the successful delivery probability of shared messages under strict resource limits. 
This is crucial in IoT applications like critical monitoring; for instance, when multiple sensors detect an industrial fault, the notification must reach its destination at least once. 
In these cases, traditional throughput-maximizing approaches are inefficient, as they assume each packet of individual node carries independent information; instead nodes must coordinate implicitly to guarantee at least one delivery per shared message. 
Achieving this goal becomes challenging in dynamic environments where the set of nodes transmitting messages changes over time. 
Dynamic node activation patterns, which evolve probabilistically from one time slot to the next, result in constantly shifting active sets. This unpredictability is compounded by the fact that activation probabilities are neither known in advance nor static; they may depend on conditional relationships between nodes or emerge randomly.
For example, a node may be more likely to be active if another correlated node is also active, or activations of two nodes may be independent of each other. Crucially, each node only knows its own activation status at any given moment, making real-time coordination with other nodes challenging. 

We also address \ac{IoT} networks operating in 
continuously evolving environments, where new sources of messages may emerge, nodes may move, and the phenomena being monitored could change \cite{drift2}. % drift1
This requires the communication protocol adapting to time-varying activation probabilities and handling the inherent uncertainty in node activations. 
These challenges, i.e., the lack of global information and the dynamics of message transmission scenarios, 
make it essential 
to develop efficient and decentralized access control mechanisms that ensure reliable and timely message delivery, while allowing to tune transmission strategies in accordance with changes in node activation patterns. 
These scenarios are especially relevant for \ac{URLLC} and energy-constrained IoT deployments \cite{urllc}, where strict latency and power budgets rule out costly retransmissions and prolonged coordination. 
\blue{While classical collision-avoidance mechanisms, such as the request-to-send/clear-to-send (RTS/CTS) handshake \cite{05-10624788}, are 
effective in standard contention-dominated networks, they are not applicable to our scenario. First, RTS/CTS is inherently a grant-based protocol (where the CTS acts as the channel grant), which directly violates our strict grant-free transmission constraint. Second, simultaneous attempts by multiple nodes to deliver a shared message would merely shift the collisions from the data phase to the RTS phase. Consequently, resolving these RTS collisions would still require the implicit coordination mechanisms we propose herein.
% {because}: (i) simultaneous attempts to deliver a shared message would merely shift collisions to the RTS phase, and (ii) multi-round handshake would violate the problem's grant-free transmission constraint.
}

\subsection{Related Work}
Several works in the literature address scenarios that share similarities with the one proposed in this paper. In \cite{unsourcedMA}, the authors introduce a framework for critical \ac{IoT} applications by prioritizing \textit{alarm messages} over \textit{standard messages} for all nodes. They assume that the number of active nodes is random and unknown in each time slot, deriving a random-coding achievability bound on the misdetection and false positive probabilities of both message types over a Gaussian multiple access channel. Although their focus is on message type classification, the random nature of node activation aligns with some aspects of our problem. In \cite{activityDetection}, the authors focus on improving detection performance in grant-free random access systems where node activation patterns are correlated. This correlation aids to enhance system performance, but the work primarily emphasizes activity detection rather than message delivery coordination 
which are central to our approach. 

\blue{To meet strict \ac{URLLC} constraints and reduce signaling overhead, recent research increasingly favors learning-based methods utilizing implicit inter-node coordination. Table \ref{tab:literature_comparison} summarizes these prior works to position our approach.} 
The work in \cite{learningToSpeak} is the closest one to the current paper. The authors consider a scenario where a single shared message must be sent by a set of nodes over shared resources, with signaling constraints necessitating a random access scheme. The nodes must coordinate implicitly without inter-node communication to ensure that at least one transmission avoids a collision. They propose a solution based on a distributed version of Thompson sampling for Bernoulli \ac{MAB} algorithms that can operate independently on each node after an initial training phase. While effective, their approach does not fully address scalability concerns as the number of nodes grows. Our work differs in that we consider the transmission of multiple shared messages and explicitly address the need to adapt to dynamic activation probabilities that govern the active sets and can continuously evolve. Other works, such as \cite{kalor2018random} and \cite{Ali2018SleepingMB}, focus on throughput maximization by leveraging correlations in activation patterns. Similarly, in \cite{StochasticResource}, the authors propose a resource allocation algorithm to design a random access scheme that maximizes both throughput and sum-rate by leveraging correlated activation patterns. These works are focused on throughput optimization, which contrasts with our goal of improving the reliability and success probability of transmitting shared messages. 

\blue{\Ac{RL}}-based techniques have recently 
been explored in throughput maximization tasks. In \cite{marl}, \ac{MARL} is used to design grant-free random access schemes suited for \ac{mMTC} scenarios. Similarly, in \cite{rl2}, \ac{RL} is employed to develop a coordinated random access scheme for industrial \ac{IoT} networks, where nodes generate sporadic, correlated traffic. \blue{Advanced frameworks, such as NOMA-PPO \cite{10-10621640}, have been developed to guarantee strict \ac{URLLC} deadlines via deep \ac{RL}. Furthermore, \ac{RL} architectures have been leveraged to achieve generalization across heterogeneous networks for \ac{MAC} protocol design \cite{01-10000805,03-10960644}}. 
However, both \ac{RL} and \ac{MARL} approaches tend to suffer from scalability issues, especially in large-scale random access schemes\blue{, and are typically bound to the paradigm of \textit{independent} message transmission, where unique resources must be allocated to unique payloads}. 

A more lightweight approach is proposed in \cite{estimationError}, where the authors develop a stochastic gradient descent-based algorithm to optimize slot allocation probabilities. While promising, their method assumes knowledge of activity estimation errors, which may not always be available in real-world settings. 
The work in \cite{GNNUplinkScheduling} propose a \ac{GNN}-based, distributed framework to optimize uplink scheduling requests in industrial \ac{IoT} scenarios, reducing redundant scheduling messages and communication overhead; unlike our work, which targets reliable transmission of multiple shared messages, their focus is on scheduling-request reduction and efficient distributed coordination.
Finally, a related problem is explored in \cite{stern2019massive}, where the authors aim to reliably transmit a shared alarm message by superposing individual signals. Unlike the collision model we adopted, this approach does not require coordination between the nodes and relies on signal superposition for reliable transmission.

\begin{table}[t]
    % \scriptsize
    \centering
    \caption{\blue{Comparison of learning-based MAC designs leveraging implicit inter-node coordination.}}
    \label{tab:literature_comparison}
    \begin{tabular}{c c c c c}
        \hline
        \textbf{Ref.} & \makecell{\textbf{Message}\\ \textbf{type}} & \textbf{Retransm.} & \makecell{\textbf{Decentralized}\\ \textbf{execution}} & \makecell{\textbf{Largest}\\\textbf{network}\\ \textbf{size}} \\
        \hline
        \cite{learningToSpeak} & Single shared & No & Yes & 20 \\
         \hline
         \cite{Ali2018SleepingMB} & Independent & Yes & No & 50 \\
         \hline
        \cite{StochasticResource} & \makecell{{Independent} / \\ {correlated}}  & Yes & No & 50 \\
        \hline
        \cite{marl} & \makecell{{Independent} / \\ {correlated}} & Yes & Yes & 20 \\
        \hline
        \cite{rl2} & \makecell{{Independent} / \\ {correlated}} & Yes & Yes & 20 \\ 
        \hline
        % Below there are new papers
        \cite{10-10621640} & Independent & Yes & No & 40 \\
        \hline
        \cite{01-10000805} & Independent & Yes & Yes & 10 \\
        \hline
        \cite{03-10960644} & Independent & Yes & Yes & 25 \\
        \hline
        % Above there are new papers
        \cite{estimationError} & Independent & No & Yes & 20 \\
        \hline
        \cite{stern2019massive} & \makecell{{Independent} / \\ {single shared}} & No & Yes & 20 \\
        \hline
        \makecell{\textbf{This} \\ \textbf{work}} & \makecell{\textbf{Multiple} \\ \textbf{shared}} & \textbf{No} & \textbf{Yes} & \textbf{64} \\
        \hline
    \end{tabular}
\end{table}

In this work, we extend the study of decentralized \ac{MAC} scenarios by \blue{considering} the challenge of transmitting multiple shared messages under dynamic activation probabilities.
\blue{By maintaining low complexity and high scalability, our approach fills a critical gap left by previous methods: {ensuring the reliable delivery of shared state information, rather than merely maximizing throughput for independent messages.}
}

\subsection{Our Contributions}

In this work, we address a novel \ac{MAC} problem where multiple shared messages must be transmitted by nodes over shared communication resources. Due to strict constraints on signaling overhead, nodes are required to rely on a local access scheme with  
no centralized coordination or explicit communication among nodes. 
The main challenge is to guarantee every shared message delivered at least once, which requires nodes 
implicitly coordinating their transmissions in a fully decentralized manner \deniz{during inference}, under dynamic node activation patterns and without the possibility of retransmissions. 
More in detail, our contributions can be summarized as follows:

\begin{enumerate} 
    \item We propose a new \ac{MAC} problem for coordinated transmission of multiple shared messages over limited resources. Analyzing its theoretical foundations, we prove that \deniz{an optimal deterministic protocol always exists, and that any randomized optimal solution must place its support on the set of} 
    optimal deterministic strategies. Moreover, we show that this problem is NP-hard. 

    \item We develop an unsupervised and stateless learning-based method 
    to solve the proposed problem. Our method is lightweight and decentralized, where each node deploys a local \ac{DNN} to select transmission opportunities. It eliminates the need of inter-node communication and central coordination, significantly simplifying the protocol design and system requirements. 

    \item We derive a theoretical upper bound on the degradation of transmission success probability when node activation probabilities change due to environment variations,  
    indicating resilience of our solution to mild environment changes. When these changes are significant, 
    we introduce an online learning mechanism 
    to maintain solution robustness to time-varying environments. 

    \item We perform extensive experiments to show that our framework outperforms baselines, scales to large networks, and exhibits robustness to dynamic environment changes. 
    All code 
    is available on GitHub\footnote{Code available at: \url{https://github.com/Lostefra/Learning-Dec-MAC}.}.
\end{enumerate}

\subsection{\blue{Notation and Organization}}
\label{subsec:notation}

\blue{To facilitate the reading of this paper and consolidate the mathematical framework, we summarize the key notations used throughout this work in Table \ref{tab:notation}. The proposed decentralized \ac{MAC} problem introduces a complex coordination challenge governed by the following core system assumptions:
\begin{itemize}
    \item Nodes have no awareness regarding the composition of the active set $\mathcal{A}_l$. They cannot communicate with one another prior to transmission, precluding the use of explicit coordination mechanisms.
    \item Transmissions must occur without centralized scheduling, multi-round handshakes (e.g., RTS/CTS), or retransmissions. This strict resource limitation is designed to conform to the tight latency and power budgets typical of URLLC deployments.
    \item During training and online adaptation, nodes receive only a single system-wide scalar reward ($\xi$) broadcasted by a central controller. During inference, nodes execute their transmission decisions entirely locally and deterministically. 
\end{itemize}
}

\begin{table}[t]
    \centering
    \caption{\blue{Summary of key notations}}
    \label{tab:notation}
    \renewcommand{\arraystretch}{1.2}
    \begin{tabular}{l p{0.75\linewidth}}
        \hline
        \textbf{Symbol} & \textbf{Description} \\
        \hline
        \multicolumn{2}{l}{\textit{Network \& Sets}} \\
        \hline
        $N$, $\mathcal{N}$ & Total number of nodes, and the set of all nodes $\{n_1,\dots,n_N\}$ \\
        $L$, $\mathcal{L}$ & Total number of shared messages, and the set of messages $\{l_1,\dots,l_L\}$ \\
        $M$ & Number of orthogonal transmission opportunities \\
        $S$ & Central controller receiving the transmissions \\
        $t$ & Discrete time slot or step \\
        $\mathcal{A}_l$ & Subset of active nodes holding message $l$ \\
        $\mathcal{A}$ & Concatenation of all active sets, $\mathcal{A}_1 \mathbin\Vert \ldots \mathbin\Vert \mathcal{A}_L$ \\
        $K$ & Maximum possible cardinality of $\mathcal{A}$ ($K = L \cdot N$) \\
        $\mathcal{P}_i$ & $i$-th possible active set pattern \\
        $\mathcal{R}_{\mathcal{A}}$ & Set of all possible subsets of active nodes, $\{\mathcal{P}_1, \ldots, \mathcal{P}_{2^K}\}$ \\
        \hline
        \multicolumn{2}{l}{\textit{Probabilities \& Distributions}} \\
        \hline
        $p(\mathcal{A})$ & Probability mass function (PMF) of the active node set $\mathcal{A}$ \\
        $p_i$ & Probability of the specific active set pattern $\mathcal{P}_i$ \\
        $\Delta^{2^K}$ & Standard $(2^K - 1)$-dimensional probability simplex \\
        $\alpha_i$ & Concentration parameters for the Dirichlet distribution \\
        \hline
        \multicolumn{2}{l}{\textit{Strategies \& Actions}} \\
        \hline
        $\mathbf{x}_{l,n}$ & Transmission move vector of node $n$ for message $l$, $\mathbf{x}_{l,n} \in \{0, 1\}^M$ \\
        $x_{l,n,m}$ & Binary indicator evaluating to 1 if node $n$ transmits $l$ over opportunity $m$ \\
        $\mathbf{X}$ & Tensor of aggregate moves of all nodes/messages, $\mathbf{X} \in \{0, 1\}^{L \times N \times M}$ \\
        $\xi_a(l,m)$ & Indicator condition: exactly one active node transmits message $l$ on opportunity $m$ \\
        $\xi_b(l,m)$ & Indicator condition: no node transmits any other message on opportunity $m$ \\
        $\xi(\mathcal{A}, \mathbf{X})$ & Overall transmission success indicator \\
        $\Phi_{l,n}$ & Local strategy (random variable) of node $n$ for message $l$ \\
        $\bm{\phi}_{l,n}$ & PMF of the local strategy $\Phi_{l,n}$ \\
        $\mathbf{\Psi}$ & Joint strategy (random variable) of all nodes \\
        $\bm{\psi}(\mathbf{X})$ & PMF of the joint strategy $\mathbf{\Psi}$ \\
        $\mathbb{E}[\,\xi \mid \mathbf{\Psi}, p\,]$ & Expected probability of successful transmission (objective function) \\
        \hline
        \multicolumn{2}{l}{\textit{Learning Framework}} \\
        \hline
        $\pi_{l,n}$ & Neural network policy deployed by node $n$ for message $l$ \\
        $\boldsymbol{\theta}_{l,n}$ & Trainable parameters of the policy $\pi_{l,n}$ \\
        $\lambda$ & Learning rate for policy gradient updates \\
        $\varepsilon$ & Exploration decay factor controlling training stochasticity \\
        $s$ & Random exploration variable sampled from a Gaussian distribution \\
        \hline
    \end{tabular}
\end{table}

The remainder of this paper is structured as follows. Sec.~\ref{sec:system_model} formulates the problem setting  
and Sec.~\ref{sec:th_find} presents theoretical findings. In Sec.~\ref{sec:learning}, we detail our proposed unsupervised learning framework and online learning mechanism, which are then evaluated through numerical simulations in Sec.~\ref{sec:numerical_results}. Finally, conclusions are drawn in Sec.~\ref{sec:conclusions}.

\section{Problem Formulation}
\label{sec:system_model}

\begin{figure}[t]
    \centering
    \includegraphics[trim= {0 0 0 0}, clip, width=0.91\columnwidth]{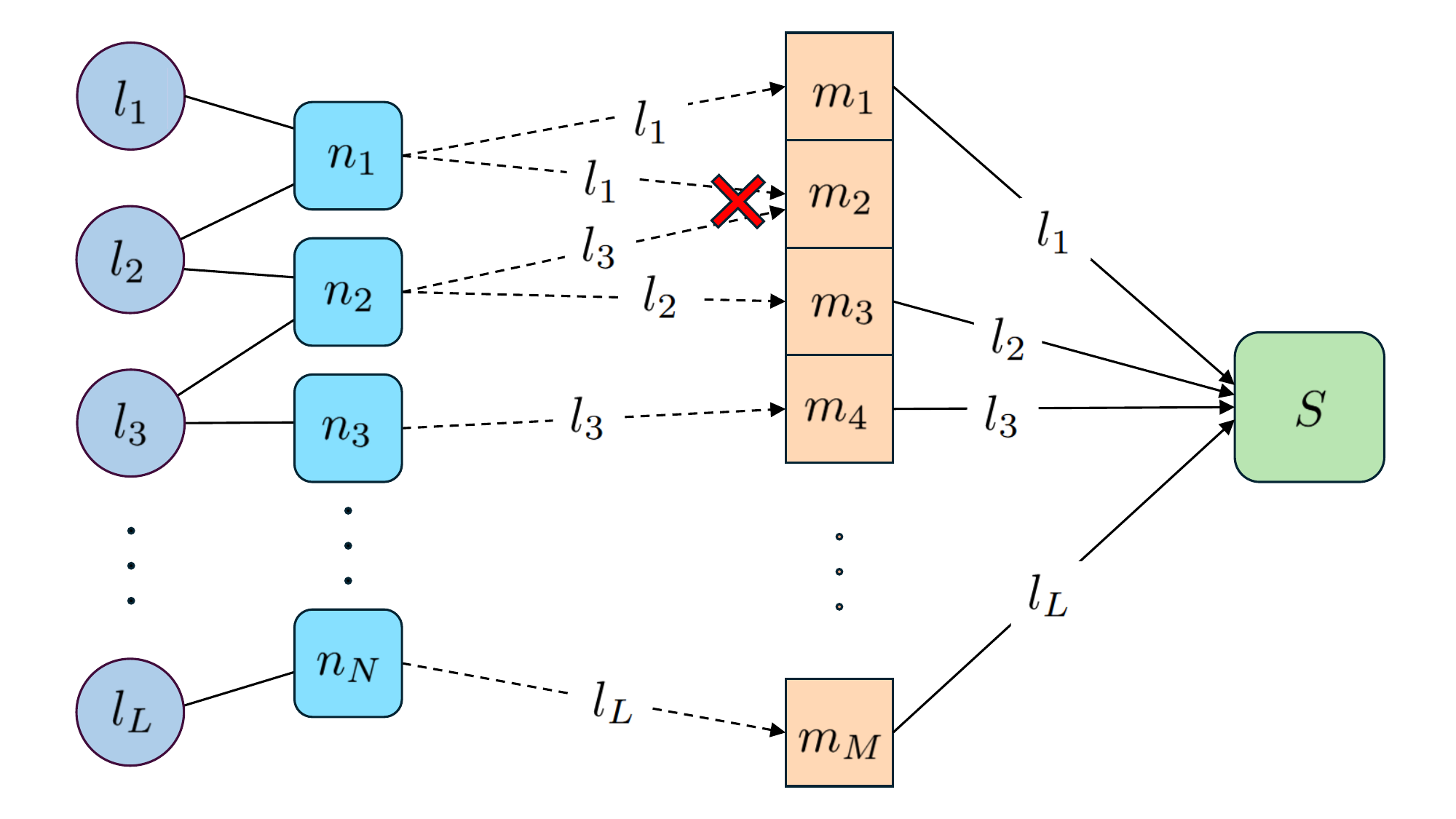}
    \caption{Reference scenario comprising $N$ nodes transmitting $L$ messages over $M$ communication opportunities to a central controller $S$. Nodes $n_1$ and $n_2$ transmit over the same communication opportunity $m_2$ simultaneously, resulting in a ``collision''. 
    }
    \label{fig:scenario}
\end{figure}

We consider a wireless network consisting of $N$ nodes $\mathcal{N}=\{n_1,\ldots,n_N\}$, which are tasked 
to transmit a set of $L$ shared messages $\mathcal{L} = \{l_1\ldots,l_L\}$ to a central controller $S$ using $M \geq L$ orthogonal transmission opportunities (in time, frequency, etc.).
At any given time $t$,\footnote{For simplicity, we omit the subscript $t$ in following expressions to improve readability.} message $l \in \mathcal{L}$ becomes available to a \deniz{non-empty} subset of \textit{active nodes}, denoted as $\mathcal{A}_l$, where $\mathcal{A}_l \subseteq \mathcal{N}$ and $0 < \vert \mathcal{A}_l \vert \leq N$. 
\blue{As introduced in Section \ref{sec:introduction}, the core objective is to ensure each message $l \in \mathcal{L}$ is successfully delivered to $S$ at least once over the $M$ opportunities. This must be achieved despite the complete absence of inter-node awareness regarding the active set $\mathcal{A}_l$, precluding explicit coordination.} 
An overall depiction of the scenario is provided in Fig.~\ref{fig:scenario}.

The coordination problem highly depends on the pattern according to which the nodes become active. 
We 
model all active nodes at time $t$ as $\mathcal{A} = \mathcal{A}_1 \mathbin\Vert \mathcal{A}_2 \mathbin\Vert \ldots \mathbin\Vert \mathcal{A}_L$, where \( \cdot\mathbin\Vert\cdot \) is the concatenation operator. 
Here, a node that is active for $\ell$ messages simultaneously is counted $\ell$ times. For the cardinality of $A$, we have $\vert \mathcal{A} \vert \leq L\cdot N \deniz{\triangleq} K$. 
We model $\mathcal{A}$ as a random variable with probability mass function $p(\mathcal{A})$. 
Note that 
$\mathcal{A}$ is sampled 
from all possible subsets of active nodes, i.e., $\mathcal{A} \in \mathcal{R}_{\mathcal{A}} = \{\mathcal{P}_1, \ldots, \mathcal{P}_{2^{K}}\}$, where the probability of the set $\mathcal{P}_{i}$ is $p_i$ for $i=1,\ldots,2^K$.  
To account for possible correlations among nodes that are active for 
the same message (e.g., device groups that tend to co-activate), we allow \(p(\mathcal{A})\) to exhibit certain correlation structure. 
\blue{In practical wireless systems, device activations are rarely uniformly random. Instead, they are often driven by shared physical events or application-specific traffic demands that cause particular subsets of nodes to co-activate. While such traffic correlations in random access systems are traditionally modeled using Markov processes, Ising models, or copulas, Dirichlet-based frameworks have emerged as a powerful tool to capture complex spatio-temporal dependencies \cite{grazian2025spatio}. From a statistical perspective, because the active set $\mathcal{A}$ follows a categorical distribution over the $2^K$ possible patterns, treating its PMF as a random variable drawn from a Dirichlet distribution is the most theoretically robust modeling choice, as it is the natural conjugate prior \cite{gelman1995bayesian}.} 
A convenient and flexible way to encode this structure is \blue{thus} to place a Dirichlet prior on the probability vector over the \(2^{K}\) possible activation patterns. 
\blue{This provides the necessary degrees of freedom to capture diverse traffic behaviors.} 
By definition, these probabilities belong to the standard $(2^{K} - 1)$-dimensional simplex 
\begin{equation}\label{eq:possibleSet}
    \Delta^{2^{K}} = \left\{ (p_1, \ldots, p_{2^{K}}) : \sum_{i=1}^{2^{K}} p_i = 1,~ p_i \geq 0 \right\}\,.
\end{equation} 
Since \blue{the PMF $(p_1, \ldots, p_{2^{K}})$ lies in} $\Delta^{2^K}$, \blue{we model it} as a sample \blue{from} a Dirichlet distribution over the simplex:
\begin{equation}\label{eq:Dirich}
    (p_1, \ldots, p_{2^K}) \sim \text{Dir}(\alpha_1, \ldots, \alpha_{2^K})~,
\end{equation}
where $\alpha_1, \ldots, \alpha_{2^K}$ are the concentration parameters with each $\alpha_i > 0$. 
These parameters determine the shape of the Dirichlet distribution. If 
$\alpha_i > 1$, the distribution is more concentrated around the center of the $(2^{K} - 1)$-dimensional simplex. If 
$\alpha_i < 1$, the distribution tends to favor extreme values, inducing a strong preference for a few recurring patterns. When all parameters are $\alpha_i = 1$ for $i=1,\ldots,2^K$, the Dirichlet distribution becomes the uniform distribution over the simplex. 

Each active node $n \in \mathcal{A}_l$ needs to decide which communication opportunities it uses to transmit message $l$. It is allowed to use multiple communication opportunities to transmit the same message, the principle behind which is 
to leverage repetitions to improve throughput and reliability, as in \ac{IRSA} schemes \cite{irsa}. We model this decision as a vector $\mathbf{x}_{l,n} \in \{0, 1\}^{M}$, referred to as a \textit{move}, where $[\mathbf{x}_{l,n}]_m = x_{l, n, m} = 1$ if node $n$ transmits message $l$ over the $m$-th opportunity. We represent the moves of all nodes for all 
messages in a tensor $\mathbf{X} \in \{0, 1\}^{L \times N \times M}$, hence we have a total of $2^{LNM}$ different possible moves. We say a transmission to be successful if the central controller receives all messages $\mathcal{L}$. Furthermore, a message is received successfully through the $m$-th opportunity only if at most one node transmits over communication opportunity $m$. 
\blue{In this context, we model the condition for a successful transmission by first defining two auxiliary indicator variables. Let $\xi_a(l,m)$ denote the condition that exactly one active node transmits message $l$ over opportunity $m$:
\begin{equation}
      \xi_a(l,m) = I \Bigg(\sum_{n \in \mathcal{A}_l} x_{l, n, m} = 1  \Bigg)~,
\end{equation}
where $I(\cdot)$ denotes the indicator function, which returns 1 if the condition is met and 0 otherwise. Similarly, let $\xi_b(l,m)$ denote the condition that no node transmits any other message on the same opportunity $m$:
\begin{equation}
      \xi_b(l,m) = I \Bigg(\sum_{\substack{\bar{l} = 1 \\ \bar{l} \neq l}}^{L} \sum_{\bar{n} \in \mathcal{A}_{\bar{l}}} x_{\bar{l}, \bar{n}, m} = 0  \Bigg)~.
\end{equation}
The overall transmission is successful if, for every message $l \in \mathcal{L}$, there exists at least one communication opportunity $m$ that satisfies both conditions. We model this overall success indicator as:
\begin{align}
\begin{split}
\label{eq:xi}
    \xi(\mathcal{A}, \mathbf{X}) &= \bigwedge_{l=1}^{L} I \Bigg( \exists m \in \{i\}_{i=1}^M :
    \xi_a(l,m) \land~\xi_b(l,m) \Bigg)~,
\end{split}
\end{align}
where \( \land \) represents the \textit{logical and} operator and $\bigwedge$ denotes the logical conjunction over multiple elements.}

For example in Fig.~\ref{fig:scenario}, consider nodes $n_1,n_2,\dots,n_N$ with message availabilities
$(l_1,l_2),(l_2,l_3),(l_3),\dots,(l_L)$, respectively. 
Node $n_1$ sends message $l_1$ on communication opportunities $m_1$ and $m_2$; $n_2$ sends $l_2$ on $m_3$ and $l_3$ on $m_2$; $n_3$ sends $l_3$ on $m_4$; $n_N$ sends $l_L$ on $m_M$. The transmission is successful, i.e., the success indicator $\xi(\mathcal{A},\mathbf{X})$ in \eqref{eq:xi} is equal to 1, although $n_1$ and $n_2$ both transmit on $m_2$ resulting in a collision. 
\blue{This is because for every shared message $l$ there exists at least one transmission opportunity $m$ satisfying the two required conditions: $\xi_a(l,m) = 1$ and $\xi_b(l,m) = 1$.} 
Concretely, $l_1$ is received via $m_1$ (only $n_1$ transmits $l_1$ on $m_1$), $l_2$ is received via $m_3$ (only $n_2$ transmits $l_2$ on $m_3$), $l_3$ is received via $m_4$ (only $n_3$ transmits $l_3$ on $m_4$), and so on up to $l_L$ on $m_M$. Since each $l\in\mathcal{L}$ has at least one interference-free opportunity, the conjunction over messages in \eqref{eq:xi} evaluates to 1 and the collective transmission is successful.

We define a \textit{local strategy} of node $n$ to 
transmit message $l$ as a random variable ${\Phi}_{l, n}$, whose support is the set of all possible moves, and its \ac{PMF} $\bm{\phi}_{l, n}(\mathbf{x}_{l,n}) \in [0, 1]$ represents the probability that node $n$ chooses move $\mathbf{x}_{l,n} \in \{0, 1\}^{M}$ when it is active to transmit message $l$. Moreover, we define the \textit{joint strategy} of all nodes as a random variable $\mathbf{\Psi}$ whose \ac{PMF} $\mathbf{\psi}(\mathbf{X}) \in [0, 1]$ is the joint \ac{PMF} of $\{\Phi_{l,n}\}_{l=1, n=1}^{L, N}$, i.e., 
\begin{align}
\begin{split}
\label{eq:pmf_prod}
    \mathbf{\psi}(\mathbf{X}) = P\left(\mathbf{\Psi} = \mathbf{X}\right) = \prod_{l=1}^{L} \prod_{n=1}^{N} \bm{\phi}_{l, n}(\mathbf{x}_{l,n})~,
\end{split}
\end{align}
where \(P(\cdot)\) denotes a probability. We can then get the expected probability of successful transmission according to the law of total probability as 
\begin{align}\label{eq:objective}
\begin{split}
    \mathbb{E}[\,\xi \mid \mathbf{\Psi}, p\,] = \sum_{\mathcal{A} \in \mathcal{R}_{\mathcal{A}}} p(\mathcal{A}) \!\!\!\!\!\!\!\!\!\!\sum_{\mathbf{X} \in \{0, 1\}^{L \times N \times M}} \!\!\!\!\!\!\!\!\!\!\xi(\mathcal{A}, \mathbf{X})~ \mathbf{\psi}(\mathbf{X})~,
\end{split}
\end{align} 
where $\mathbb{E}[\cdot]$ is the expectation, and 
formulate an optimization problem to maximize the probability of successful transmission by finding 
the optimal joint strategy $\mathbf{\Psi}^*$ as
\begin{align}
\begin{split}
\label{eq:optim}
   \mathbb{E}[\,\xi \mid \mathbf{\Psi}^*, p\,] =~\maxb_{\mathbf{\Psi}}~\mathbb{E}[\,\xi \mid \mathbf{\Psi}, p\,] ~,
\end{split}
\end{align}
referred to as the problem of shared message transmission. 
\blue{We emphasize that while the performance metric $\xi$ and the objective function in \eqref{eq:optim} are evaluated from a system-wide perspective, the generation of the transmission moves $\mathbf{X}$ is fundamentally decentralized. The joint strategy $\mathbf{\Psi}$ explicitly factors into independent local strategies $\Phi_{l,n}$, meaning nodes execute their transmission decisions without explicit coordination or state-sharing. This theoretical separation of a global objective and local decision-making naturally motivates the \ac{CTDE} framework proposed in Section \ref{sec:learning}.} 
In Section \ref{sec:th_find}, we conduct theoretical analysis to provide interpretations for problem \eqref{eq:optim} and its optimal solution. In Section \ref{sec:learning}, we 
develop a \blue{\ac{CTDE}-based} method to solve problem \eqref{eq:optim} and leverage online learning that adapts our method to dynamic wireless scenarios. In Section \ref{sec:numerical_results}, we evaluate the proposed method in extensive numerical experiments to show superior performance. 

\section{Theoretical Analysis}
\label{sec:th_find}

In the problem of shared message transmission \eqref{eq:optim}, 
understanding the behaviors and properties 
of the optimal joint strategies 
is critical, as it provides valuable insights to efficiently guide the search of 
solutions. Before proceeding, we define deterministic and randomized joint strategies of problem \eqref{eq:optim} as follows. 
\begin{definition}[Deterministic joint strategy] 
A joint strategy $\mathbf{\Psi}$ 
is said to be \emph{deterministic} if its \ac{PMF} takes binary values. That is, for every node $n\in\mathcal{N}$ and for every message $l\in\mathcal{L}$, there exists a unique move $\mathbf{x}_{l,n}^{*} \in \{0,1\}^{M}$ such that
\begin{align}
    \bm{\phi}_{l,n}(\mathbf{x}_{l,n}^{*}) &= 1, \\
    \bm{\phi}_{l,n}(\mathbf{x}_{l,n}) &= 0, \quad \forall\, \mathbf{x}_{l,n} \neq \mathbf{x}_{l,n}^{*}\,.
\end{align}
Equivalently, the joint PMF $\mathbf{\psi}(\mathbf{X})$ satisfies
\begin{align}
    \mathbf{\psi}(\mathbf{X}) \in \{0,1\},\quad \forall~\mathbf{X}\in\{0,1\}^{L \times N \times M}\,,
\end{align}
implying that the strategy always selects the same moves.
\end{definition}
\begin{definition}[Randomized joint strategy]
A joint strategy $\mathbf{\Psi}$ 
is said to be \emph{randomized} if there exists at least one node $n\in\mathcal{N}$ and one message $l\in\mathcal{L}$ for which the corresponding PMF $\bm{\phi}_{l,n}$ is non binary-valued. That is, there exists at least one move $\mathbf{x}_{l,n} \in \{0,1\}^{M}$ such that
\begin{align}
    0 < \bm{\phi}_{l,n}(\mathbf{x}_{l,n}) < 1\,.
\end{align}
That is, the joint strategy $\mathbf{\Psi}$ inherently involves probabilistic decision-making.
\end{definition}
One of the 
key questions is whether deterministic strategies are sufficient to achieve optimality or randomization is needed.
Deterministic strategies are easier to interpret and provide 
predictable behavior compared to randomized strategies, as they represent a specific subset within the broader class of randomized strategies and each node 
makes always the same move given the same problem setting. This motivates us to characterize the deterministic / randomized property of the optimal strategies for problem \eqref{eq:optim} in the following theorem. 
\begin{theorem}
\label{th:1}
\textit{The set of optimal joint strategies for the optimization problem defined in (\ref{eq:optim}) either consists of a single deterministic joint strategy or includes a finite number $D > 1$ of deterministic joint strategies along with infinitely many randomized ones, i.e.,} $\forall\,\mathbf{X} \in \{0, 1\}^{L \times N \times M}$:
\begin{align}
\begin{gathered}
   \exists!~\mathbf{\Psi},~\mathbf{\psi}(\mathbf{X}) \in \{0, 1\} : \mathbb{E}[\,\xi \mid \mathbf{\Psi}\,] = \mathbb{E}[\,\xi \mid \mathbf{\Psi}^*\,] \\
   \oplus \\
   \exists~\{\mathbf{\Psi}_i' \}^D_{i=1},~\mathbf{\psi}_i'(\mathbf{X}) \in \{0, 1\}, D > 1~\land \\
   \exists~\{\mathbf{\Psi}_j'' \}^\infty_{j=1},~\mathbf{\psi}_j''(\mathbf{X}) \in [0, 1] : \\
   \mathbb{E}[\,\xi \mid \mathbf{\Psi}_i'\,] = \mathbb{E}[\,\xi \mid \mathbf{\Psi}_j''\,] = \mathbb{E}[\,\xi \mid \mathbf{\Psi}^*\,]~,
\end{gathered}
\end{align}
where $\oplus$ is the \textit{exclusive or}, the existence of an element $x$ is denoted by \( \exists x \), while \( \exists! x \) specifies that $x$ exists uniquely.
\end{theorem}
\begin{proof}
    See Appendix \ref{appendix:A}.
\end{proof}
Theorem \ref{th:1} reveals a fundamental property of the optimal joint strategies for problem \eqref{eq:optim}. That is, the optimality landscape is structured in one of two distinct ways: (i) 
it is possible that there exists a unique deterministic joint strategy that achieves the maximum expected performance; (ii) 
if there are multiple optimal deterministic joint strategies (i.e., a finite set of them), 
there exists an infinite number of randomized strategies - constructed as convex combinations of those deterministic strategies - that also achieve the same optimal performance.
% Therefore, 
% it suffices to focus only on deterministic joint strategies when solving problem \eqref{eq:optim} 
% without loss of optimality. 
% This result motivates to search for deterministic joint strategies, which offer a clear and unambiguous rule for decision-making, without resorting to probabilistic rules. 
%
\deniz{However, recall that our system strictly enforces independent decision-making, meaning any valid decentralized joint strategy must factorize into a product distribution of the local strategies [cf. \eqref{eq:pmf_prod}]. Due to the complete lack of inter-node communication, nodes cannot explicitly coordinate to execute arbitrary joint randomizations or synchronously switch between multiple optimal deterministic strategies. If independent local randomizations were used to approximate a convex combination of optimal deterministic strategies, it would inevitably lead to mismatched moves and heavily degraded performance. Therefore, to safely achieve optimal performance without requiring inter-node coordination, the nodes must implicitly align on a \textit{single} deterministic strategy.} This result motivates to search for deterministic joint strategies, which offer a clear and unambiguous rule for decision-making, without resorting to probabilistic rules. 

While Theorem \ref{th:1} provides interpretations for the optimal solution and narrows down the search space of transmission strategies, problem \eqref{eq:optim} remains significantly difficult 
to solve. 
In the following theorem, we formally show that \eqref{eq:optim} is a NP-hard problem. 

\begin{theorem}\label{thm2}
\textit{The optimization problem in (\ref{eq:optim}) is NP-hard.}
\end{theorem}
\begin{proof}
    See Appendix \ref{appendix:A}.
\end{proof}

Theorem \ref{thm2} states that it is indeed challenging to solve problem \eqref{eq:optim}, which corroborates our intuitive observations from its nature of limited resources, message sharing and local information availability. The result identifies the complexity of conventional heuristic methods when solving problem \eqref{eq:optim} and motivates to leverage learning-based approaches to develop efficient and scalable solutions in a decentralized manner. 

In general, it is possible that node activation probabilities vary over time due to 
many different 
factors, such as 
new sources of messages, changes in the wireless environment, node mobility, and so on. This indicates 
that the distribution of active node sets $\mathcal{A}$, modeled by $p(\mathcal{A})$, can vary over time. As a consequence, 
% an important property that requires attention %consideration 
\deniz{a natural question} 
is the robustness of a joint strategy $\mathbf{\Psi}$ to distribution \deniz{shifts in} $\mathcal{A}$, 
i.e., how much $\xi' = \mathbb{E}[\,\xi \mid \mathbf{\Psi},\,p'\,]$ can degrade if the distribution 
shifts to $p''$ and the expected probability of successful transmission becomes $\xi'' = \mathbb{E}[\,\xi \mid \mathbf{\Psi},\,p''\,]$. If we model $p'$ and $p''$ as two \ac{i.i.d.} realizations of the same Dirichlet distribution and 
consider $\mathbf{\Psi}$ to be deterministic, the following theorem 
derives an upper bound on the probability that the degradation exceeds a threshold error, i.e., $\vert \xi' - \xi'' \vert \geq \eta$ with $\eta \in [0, 1]$ the threshold error.

\begin{theorem}
\label{th:3}
\textit{Given} \deniz{i.i.d.} $p', p'' \sim \text{Dir}(\alpha_1, \ldots, \alpha_{2^K})$\textit{ [cf. \eqref{eq:Dirich}], a deterministic joint strategy $\mathbf{\Psi}$, and defining $\xi' = \mathbb{E}[\,\xi \mid \mathbf{\Psi},\,p'\,],~\xi'' = \mathbb{E}[\,\xi \mid \mathbf{\Psi},\,p''\,]$, and $\alpha_{\varepsilon} = \min \{\alpha_i \}^{2^K}_{i=1}$}, then
\begin{align}
\begin{split}
\label{eq:th3}
    \deniz{P(\vert \xi' - \xi'' \vert \geq \eta) \leq \frac{1}{2\,\eta^2\,(1 + \alpha_{\varepsilon}\,2^{K})}~,}
\end{split}
\end{align}
\textit{for any threshold $\eta \in (0, 1]$.}
\end{theorem}
\begin{proof}
    See Appendix \ref{appendix:A}.
\end{proof}

Theorem \ref{th:3} quantifies the 
\deniz{fluctuation of the success rate when $p$ is resampled from the same Dirichlet prior}.
The result indicates that when distribution shifts are mild, the transmission strategy is able to maintain satisfactory performance without need of re-design. Moreover, the bound derived in \eqref{eq:th3} is particularly useful when we consider large scenarios comprising many nodes and messages. 
\deniz{As the network grows, the Dirichlet PMF concentrates and the success-rate fluctuation between two i.i.d. realizations vanishes exponentially}. 
\deniz{However, for distribution drifts that change the prior itself, no such bound holds, and the online adaptation mechanism of Sec.~\ref{subsec:onlineLearning} is required}.
\section{MAC Protocol Learning}
\label{sec:learning}

With the deterministic nature of the optimal transmission strategy [cf. Theorem \ref{th:1}] and the NP-hard complexity of the shared message transmission problem [cf. Theorem \ref{thm2}], we propose a decentralized, unsupervised, and stateless learning-based (DUSL) algorithm to solve problem \eqref{eq:optim}. 
Firstly, DUSL is decentralized, where each node $n \in \mathcal{N}$ is associated with a distinct \ac{DNN} and determines independently its strategy to transmit messages $\mathcal{L}$ over the $M$ opportunities, without requiring communication among 
nodes. Secondly, DUSL is unsupervised because 
we update DNN parameters 
using only \acp{ACK} for successful transmissions as feedback, without relying on any oracle or ground truth. Thirdly, DUSL is stateless \deniz{in the sense that} it does not need explicit knowledge of the distribution of active node sets $p$, but instead learns to coordinate nodes' message transmission 
through trial and error during the training phase. 

\begin{figure}[t]
    \centering
    \includegraphics[trim= {0 0 0 0}, clip, width=0.91\columnwidth]{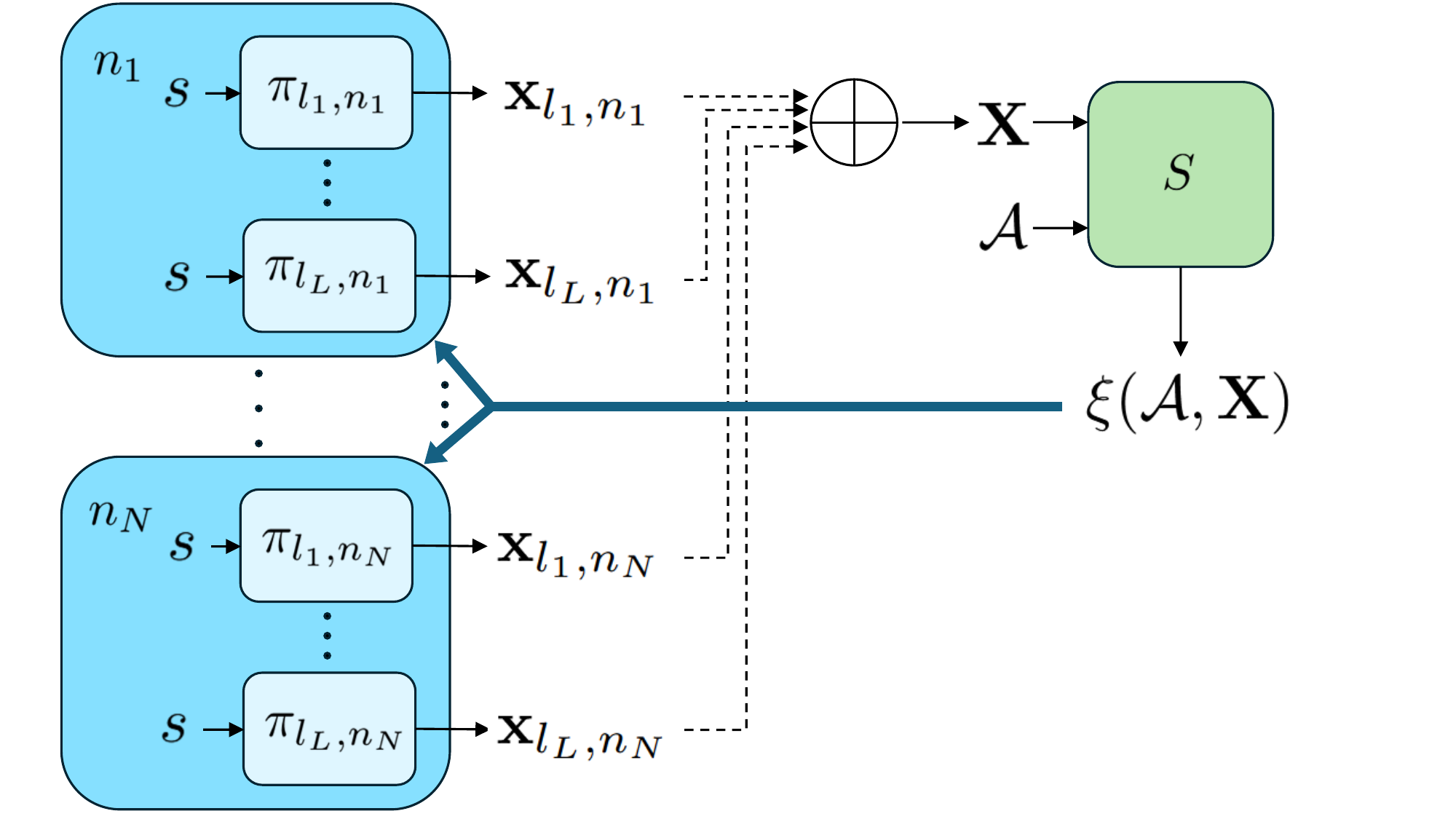}
    \caption{DUSL learning framework: Each node independently determines communication strategies for messages using $L$ local DNNs, where feedback $\xi$ guides local training procedures.
    }
    \label{fig:methodology}
\end{figure}

DUSL employs CTDE. 
At training time, we leverage a centralized feedback mechanism to guide the learning process. \blue{Specifically, the global success indicator $\xi$ defined in Section \ref{sec:system_model} is computed by the server to evaluate the joint performance of the active nodes,} and we perform 
a stochastic training procedure that enables gradient backpropagation for optimization with non-differentiable ACK feedback. At inference time, \blue{the execution is entirely decoupled from the centralized objective computation:} each node \blue{acts as an independent agent that implements its learned policy for local inference,} deploying the learned policy locally in a decentralized manner, without the operational overhead of a central controller. Moreover, the learned policy is applied in a deterministic manner \blue{(aligning with the theoretical findings in Section \ref{sec:th_find})}, where each node follows a fixed decision rule based on the trained DNN parameters to improve computational and energy efficiency, reduce inference latency, and provide interpretability. The overall training and inference procedures are presented in Algorithms \ref{alg:training}-\ref{alg:deterministic_inference} and depicted in Fig.~\ref{fig:methodology}. We provide details in the following subsections.

\subsection{DNN parameterization}

We parameterize the transmission strategy of each node $n$ as a set of \(L\) \acp{DNN} \cite{sze2017efficient, 9620104}, each corresponding to one message. For each message \(l\), the transmission decision is computed by a dedicated \ac{DNN} that consists of a single input neuron, \(H\) hidden layers, and \(M\) output neurons as
\begin{align}
    \textbf{Z}^{(l)}_n = \textbf{W}^{(l)}_{n,H} \sigma\Big(\textbf{W}^{(l)}_{n,H-1}\sigma\Big(\cdots \textbf{W}^{(l)}_{n,1} s\Big)\Big),
\end{align}
where \(\{\textbf{W}^{(l)}_{n,h}\}_{h=1}^H \in \mathbb{R}^{F_h \times F_{h-1}}\) are the linear operation matrices for the \(l\)-th \ac{DNN} of node $n$, \(F_h\) is the feature dimension at layer \(h\), and \(\sigma(\cdot)\) is the nonlinearity. The single neuron input \(s\) does not encode any node-specific state but is solely used to facilitate the exploration of new moves. 
\blue{Specifically, \(s\) acts as a random exploration variable sampled from a zero-mean Gaussian distribution during training. As the variance of this distribution decays over time, the network naturally learns to adapt to \(s=0\) as input.} 
Details are further discussed in Sec.~\ref{subsec:training}. 
The \(M\) output neurons of each \ac{DNN} are passed through a sigmoid activation function to obtain probabilities, i.e., 
\begin{align}
    [\tilde{\textbf{Z}}^{(l)}_n]_m = \sigma([\textbf{Z}^{(l)}_n]_m), \quad m = 1,\dots, M,
\end{align}
where \([\tilde{\textbf{Z}}^{(l)}_n]_m\) represents the probability that the \(l\)-th message is transmitted utilizing the \(m\)-th available transmission opportunity by node $n$.
Each \ac{DNN} can be represented as a nonlinear mapping \(\chi_{l,n}(s, \mathcal{W}^{(l)}_n)\), where \(\mathcal{W}^{(l)}_n\) collects the parameters \(\{\textbf{W}^{(l)}_{n,h}\}_{h=1}^H\) for the \(l\)-th network. To determine the transmission move of each active node $\{\mathbf{x}_{l,n}\}_{l=1}^{L}$, 
we sample from \(L \cdot M\) \deniz{independent} Bernoulli \deniz{random variables} with the corresponding probabilities \(\{[\tilde{\textbf{Z}}^{(l)}_n]_m\}_{m=1, l=1}^{M,L}\)\footnote{\deniz{If $n \notin \mathcal{A}_l$, we set } $[\tilde{\textbf{Z}}^{(l)}_n]_m=0$ for \deniz{all $m$ (i.e., a non-active node does not transmit).}}. 

The success of the \(m\)-th Bernoulli variable for the \(l\)-th \ac{DNN} indicates that the node attempts to use the \(m\)-th opportunity to transmit 
the \(l\)-th message.
We note that this parameterization does not prevent a node from selecting the same transmission opportunity for multiple messages. However, as detailed in the following section, the training procedure relies on centralized \ac{ACK} feedback and thus, DUSL will implicitly coordinate nodes and disfavor self-colliding strategies over time to increase the number of successful \acp{ACK}. Extending the \ac{DNN} 
parametrization to ensure selecting each resource at most once is left for future work.

\blue{
\textit{Remark on Parameterization:} DUSL 
uses input-free local policies: 
% operates in a strictly stateless manner, meaning the neural networks do not receive node-specific state information or identifiers as input
the only network input is the exploration scalar $s$, and no node identifier or environmental state is encoded. 
If all nodes shared a single} \blue{DNN parameterization, they would output identical probability distributions for message transmission. This symmetry would severely hinder coordination, as nodes would inherently gravitate toward the same communication opportunities, maximizing collisions. Independent DNNs $\mathcal{W}^{(l)}_n$ specific to each node $n$ naturally break this symmetry, allowing each node to learn orthogonal and asymmetric transmission strategies. 
Furthermore, independent local parameterizations eliminate the need for gradient aggregation or weight synchronization among nodes, preserving decentralization and avoiding communication overhead that would otherwise bottleneck practical IoT deployments.
}

\begin{algorithm}[t]
\footnotesize
\caption{DUSL training procedure}
\label{alg:training}
\begin{algorithmic}[1]
\State Initialize a \ac{DNN} $\pi_{l, n} \text{ with parameters }\boldsymbol{\theta}_{l, n}~\forall (l, n) \in \mathcal{L} \times \mathcal{N}$, total number of training epochs $T$, an exploration decay factor $\varepsilon \in \mathbb{R^+}$, and a learning rate $\lambda \in \mathbb{R^+}$
\For{$t \in \{1, \dots, T\}$}
    \State Sample $\bigcup_{l=1}^{L} \mathcal{A}_l = \mathcal{A}$ and $s \sim \text{Gaussian}(0, e^{-\varepsilon t})$
    \For{$l \in \mathcal{L}, n \in \mathcal{A}_l$}
        \State Get probabilities $[\tilde{\textbf{Z}}^{(l)}_n]_m \leftarrow \pi_{l,n}(s)$
        \State Sample $m_{l,n}^m \sim \text{Bernoulli}(\tilde{\textbf{Z}}^{(l)}_{n,m})$ , $m \in \{1, \cdots, M\}$
        \State Determine move $\mathbf{x}_{l,n} \in \mathbf{X} \text{ according to } m_{l, n}^1, \dots, m_{l, n}^M$
    \EndFor
    \State Compute $\xi(\mathcal{A}, \mathbf{X})$ centrally and broadcast it to $\mathcal{N}$
    \For{$l \in \mathcal{L}, n \in \mathcal{A}_l$}
        \State $\boldsymbol{\theta}_{l, n} \leftarrow \boldsymbol{\theta}_{l, n} + \lambda \cdot \nabla_{\boldsymbol{\theta}_{l, n}} \log \pi_{l, n}\left(m_{l, n}^1, \dots, m_{l, n}^M \middle| s\right) \cdot \xi(\mathcal{A}, \mathbf{X})$
    \EndFor
\EndFor
\end{algorithmic}
\end{algorithm}

\subsection{Training Procedure}
\label{subsec:training}

Our training algorithm is based on a reward-driven policy gradient approach, similar to the REINFORCE algorithm \cite{reinforce} but in an unsupervised manner and in a CTDE setting. 
At its core, the objective is to maximize the reward, i.e., expected success rate 
$\mathbb{E}[\,\xi \mid \mathbf{\Psi},\,p\,]$ [cf. \eqref{eq:objective}], over the transmission decisions taken by all active nodes. The following steps break down the procedure:

\begin{enumerate}
    \item \textit{Initialization}. For each message \(l \in \mathcal{L}\) and node \(n \in \mathcal{N}\), we initialize a neural network policy \(\pi_{l,n}\) with parameters \(\boldsymbol{\theta}_{l,n}\). 
    Each \(\pi_{l,n}\) outputs \(M\) independent probabilities corresponding to the \(M\) transmission opportunities. We set the total number of training epochs \(T\), a learning rate \(\lambda \in \mathbb{R}^+\), and an exploration decay factor \(\varepsilon \in \mathbb{R}^+\) that controls the injected randomness during training. 
    \item \textit{Exploration and action selection}. At each epoch $t$, a set of active nodes \(\mathcal{A}\) is sampled from the overall node set, and a scalar \(s\) is sampled from a Gaussian distribution \(s \sim \text{Gaussian}(0, e^{-\varepsilon t})\). 
    This noise $s$ is 
    injected to induce stochasticity in the action selection and to guide the policy exploring the action space during training. The decay factor \(e^{-\varepsilon t}\) decreases over time, which enables 
    a shift from exploration to exploitation. For each active node \(n \in \mathcal{A}_l\) corresponding to message \(l\), the policy \(\pi_{l,n}(s)\) is queried to output a set of probabilities for the \(M\) opportunities 
    \begin{equation}
        [\tilde{\textbf{Z}}^{(l)}_n]_m = \pi_{l,n}(s).
    \end{equation}
    The value of 
    $\tilde{\textbf{Z}}^{(l)}_{n,m}$ is interpreted as the probability of using the \(m\)-th transmission opportunity to transmit the \(l\)-th message. For each 
    $\tilde{\textbf{Z}}^{(l)}_{n,m}$, a Bernoulli random variable \(m_{l,n}^m\) is sampled as 
    \begin{equation}
      m_{l,n}^m \sim \text{Bernoulli}(\tilde{\textbf{Z}}^{(l)}_{n,m})
    \end{equation}
    These sampled outcomes determine the move \(\mathbf{x}_{l,n}\) of node \(n\) for message \(l\).
    \item \textit{Reward computation}. The reward signal $\xi$ is the only centralized element of the DUSL framework and is used exclusively for policy updates. Specifically, a central controller collects the messages resulting from joint transmissions, computes a \deniz{single} scalar reward $\xi$ that measures overall transmission success, and broadcasts this feedback to all $N$ nodes at each iteration. Specifically, the controller attempts to decode every communication opportunity after each joint transmission, i.e., it builds per-message ACKs $y_l\in\{0,1\}$ (1 if message $l$ was decoded on any opportunity), and broadcasts the scalar reward $\xi=\prod_{l=1}^L y_l$ for policy updates. The reward signal is positive if, and only if, the collective transmissions were successful, which indicates that every shared message was received by the controller at least once. 
    Therefore, each node learns to optimize its behavior based on system-wide performance rather than individual transmission success, promoting coordinated strategies without direct inter-node communication. We remark that $\xi$ is only needed for policy updates, % during training, 
    but not required during deployment. This ensures that the trained system operates in a completely decentralized manner, i.e., each node can execute its learned policy locally without the operational overhead of a controller.  
    \item \textit{Policy gradient update}. Each neural network policy \(\pi_{l,n}\) is updated with gradient descent in an unsupervised manner. However, since the reward \(\xi(\mathcal{A}, \mathbf{X})\) is non-differentiable [cf. \eqref{eq:xi}], the gradient cannot be computed directly. We leverage the policy gradient following 
    the REINFORCE update rule, which is a stochastic approximation for the true gradient \cite{sutton1999policy, gao2020resource, 10279331, 10622170}. 
    The update for the parameters \(\boldsymbol{\theta}_{l,n}\) is then given by:
    \begin{equation}
        \boldsymbol{\theta}_{l,n} \!\leftarrow\! \boldsymbol{\theta}_{l,n} \!+\! \lambda \cdot \nabla_{\boldsymbol{\theta}_{l,n}}\!\! \log \pi_{l,n}\big(m_{l,n}^1, \dots, m_{l,n}^M \!\mid\! s\big) \!\cdot \xi(\mathcal{A},\! \mathbf{X})
    \end{equation}
    The preceding update has three main components: (i) learning rate \(\lambda\), which controls the step size; (ii) log-probability gradient \(\nabla_{\boldsymbol{\theta}_{l,n}} \log \pi_{l,n}(\cdot)\) , which computes the sensitivity of the chosen actions with respect to the network parameters; (iii) reward weighting 
    \(\xi(\mathcal{A}, \mathbf{X})\), which reinforces actions resulting in higher rewards.  
    \item \textit{Iteration}. The process repeats over \(T\) epochs iteratively, 
    refining transmission policies 
    through observed rewards. 
    As the exploration noise $s$ decays, the \acp{DNN} focus more on exploiting the learned strategies that maximize the transmission success. 
\end{enumerate}

\subsection{Deterministic Inference}

Unlike training - where exploration via a stochastic policy is essential for learning - we switch the trained stochastic DNN-based policy to a deterministic transmission strategy 
during inference to improve 
computational and energy efficiency, reduce inference latency, and enhance interpretability. From Theorem \ref{th:1}, an optimal 
deterministic strategy \deniz{always exists; hence inference can be carried out deterministically without loss of optimality}. 
Specifically, we  
set the exploration noise 
to zero $s=0$, 
\blue{which corresponds to the mean of the training distribution of $s$. Because the optimal transmission policy is entirely encoded within the trained weights and biases of the networks, providing an input of $s=0$ produces the action distribution learned during the training phase. This} 
fixes the input of DNNs, and \blue{allows us to} precompute network outputs $\{\pi^*_{l, n}(0)\}_{l,n}$ to obtain 
decision vectors of nodes. 
Then, we round these output probabilities to the nearest integers ($0$ or $1$) to determine transmissions for all nodes and messages. This is a deterministic strategy, which remains unchanged over time and depends only on the distribution of active nodes over shared messages $p$, not instantaneous realizations of active node sets. Alg.~\ref{alg:deterministic_inference} summarizes 
the above inference procedure. 
This deterministic inference enjoys the following benefits: 
\begin{itemize}
    \item \textit{Computational and energy efficiency}. Since input remains fixed and outputs can be precomputed, the system avoids querying the neural networks repeatedly during inference, reducing unnecessary energy consumption.
    \item \textit{Reduced inference latency}. Precomputed decisions allow for immediate access to the optimal action, significantly reducing the time required for MAC decisions.
    \item \textit{Interpretability}. The deterministic decisions provide a clear and unambiguous rule for communication, facilitating easier analysis of the MAC layer.
\end{itemize}
While the training procedure requires a stochastic policy to explore and learn effective strategies via policy gradients, the inference procedure 
leverages deterministic and precomputed decisions to provide 
efficient and interpretable \ac{MAC}.

\begin{algorithm}[t]
\footnotesize
\caption{Deterministic inference procedure}
\label{alg:deterministic_inference}
\begin{algorithmic}[1]
\State \textbf{Precomputation Phase:} 
\ForAll{\(l \in \mathcal{L}\) and \(n \in \mathcal{N}\)}
    \State Compute probabilities $[\tilde{\textbf{Z}}^{(l)}_n]_m \leftarrow \pi_{l,n}(0)$
    \State Round to binary values \(m_{l, n}^1, \dots, m_{l, n}^M \leftarrow [\tilde{\textbf{Z}}^{(l)}_{n,1}], \dots, [\tilde{\textbf{Z}}^{(l)}_{n,M}]\)
    \State Compute and store move \(\mathbf{x}_{l,n} \in \mathbf{X}\) based on \(m_{l, n}^1, \dots, m_{l, n}^M\)
\EndFor
\State \textbf{Inference Phase:}
\State Sample the active node sets \(\bigcup_{l=1}^{L} \mathcal{A}_l = \mathcal{A}\)
\ForAll{\(l \in \mathcal{L}\) and \(n \in \mathcal{A}_l\)}
    \State Retrieve precomputed move \(\mathbf{x}_{l,n}\)
\EndFor
\end{algorithmic}
\end{algorithm}

\subsection{Online Learning Mechanism} 
\label{subsec:onlineLearning}

While Theorem \ref{th:3} guarantees that, under 
mild 
changes in the node activation distribution \(p\), the trained transmission strategy $\{\pi^*_{l, n}\}_{l,n}$ is capable of maintaining satisfactory performance with bounded degradation 
$\vert \xi' - \xi'' \vert$, 
practical scenarios may experience severe 
distribution shifts in \(p\). In these cases, 
$\{\pi^*_{l, n}\}_{l,n}$ could suffer from significant performance degradation. 
To overcome this issue, we propose an online learning mechanism to tune the offline trained strategy and adapt it to the shifted activation statistics in a real-time manner.  
The online learning mechanism 
is based on a reward-driven policy gradient approach as well, where the objective remains to maximize the expected success rate $\mathbb{E}[\,\xi \mid \mathbf{\Psi},\,p\,]$ over the transmission decisions taken by all active nodes. Similarly to the offline training procedure, it computes the instantaneous objective at each time step and leverages the latter to update neural networks' parameters with policy gradient, while requiring only feedback from the central controller for objective computation based on acknowledgments of successful transmissions. 
\blue{Crucially, to comply with strict URLLC latency and energy budgets, this mechanism relies on the decoupling of the \emph{real-time transmission path} (inference) from the \emph{asynchronous online tuning path} (gradient update). During the active transmission phase, nodes execute a single forward pass of their policies to make immediate MAC decisions. The gradient-based policy updates are executed asynchronously in the background upon receiving the central controller's feedback, ensuring that the critical transmission sequence is never blocked.}

Specifically, given the offline trained policies $\{\pi^*_{l, n}\}_{l,n}$ with parameters $\{\boldsymbol{\theta}^*_{l, n}\}_{l,n}$, we consider two online tuning strategies: (i) \textit{full adaptation}, which updates all parameters of $\{\pi^*_{l, n}\}_{l,n}$, 
and (ii) \textit{partial adaptation}, which updates 
only a subset of parameters while freezing the others. 
The former adapts more thoroughly to distribution shifts in time-varying environments and may yield better performance, while the latter 
reduces computational cost and speeds up the real-time online learning procedure, 
which is beneficial 
in resource-constrained settings \blue{(as detailed in Section \ref{subsec:comp_complexity})}.
For either full adaptation or partial adaptation, let $\{\pi^*_{l, n, t}\}_{l,n}$ be the transmission policies at each time step $t$ during inference. The nodes deploy their policies $\{\pi^*_{l, n, t}\}_{l,n}$ with the 
exploration variable $s$ to determine which opportunities are used and, therefore, the complete set of moves 
to transmit messages. 
Consequently, the remote server computes the transmission success rate $\xi$ 
and distributes it to nodes as feedback. Based on this feedback, each node updates full or partial parameters of its own policy $\pi^*_{l, n, t}$ with gradient descent locally through policy gradient. 

\subsection{Learning Complexity}
\label{subsec:complexity}

The 
proposed DUSL 
directly outputs $M$ probabilities - one per communication opportunity - thereby framing the problem as a structured prediction task where each binary decision determines whether to use a specific opportunity for transmission.
However, the distributed \ac{MAB} approach based on Thompson sampling presented in \cite{learningToSpeak} requires each active node to choose from \(2^M\) possible arms, each corresponding to a unique combination of the \(M\) opportunities. This exponential increase not only escalates the memory requirements but also imposes a severe burden on the learning process. 

From a learning-theoretic perspective, the hypothesis class for selecting among \(2^M\) actions entails a multi-class classification problem with an exponentially large label space because each unique combination of communication opportunities is treated as a distinct class label, 
which leads to a Vapnik–Chervonenkis (VC)-dimension \cite{blumer1989learnability} equal to $2^M$. As the VC-dimension grows, the sample complexity - the number of training examples required to achieve a desired generalization error - scales poorly. Under standard assumptions\footnote{These sample complexity estimates hold under standard learning theory assumptions \cite{shalev2014understanding}: (i) training samples are drawn i.i.d. from a fixed distribution, (ii) the loss function is bounded, (iii) the hypothesis class has finite complexity (e.g., VC-dimension), and (iv) the output structure is fixed, either as a single \(2^M\)-way output or \(M\) binary outputs. 
}, the sample complexity for learning the distributed \ac{MAB} using Thompson sampling approach scales as
\[
\mathcal{O}\left(\frac{2^M + \log(1/\delta)}{\epsilon^2}\right),
\]
where \(\epsilon\) is the target generalization error, \(\delta\) is the confidence level, and the exponential factor is due to VC-dimension given by the need to choose among \(2^M\) actions.
In contrast, by outputting \(M\) independent probabilities, the proposed DUSL effectively decomposes the task into \(M\) binary classification problems, each corresponding a binary decision. This factorization reduces 
the hypothesis class complexity, decreases its effective VC-dimension to $M$, and consequently lowers the sample complexity. This allows DUSL to reduce the sample complexity for learning to 
\[
\mathcal{O}\left(\frac{M + \log(1/\delta)}{\epsilon^2}\right),
\]
thereby providing an exponential-to-linear improvement compared to distributed MAB. 
This improvement in sample efficiency translates into faster convergence and more reliable learning performance, particularly when \(M\) is large.

\section{Numerical Results}
\label{sec:numerical_results}

We evaluate the proposed algorithm via simulations. While sampling the \ac{PMF} of active node sets $\mathcal{A}$ from a Dirichlet distribution models the general distribution $p$ (Sec.~\ref{sec:system_model}), defining concentration parameters to capture correlated activations or bound the number of active nodes remains challenging. Moreover, as the number of nodes and messages 
increases, generating the distribution $p$ completely becomes impractical because it requires defining a \ac{PMF} for $\vert \mathcal{R}_{\mathcal{A}} \vert = 2^{K}$ possible events [cf. \eqref{eq:possibleSet}]. Therefore, we define two feasible methods to determine $p$ that allow for the expression of possible pattern correlations and boundaries:

\begin{figure*}[t]
    \centering
    \begin{subfigure}[b]{0.24\linewidth}
        \includegraphics[width=\textwidth]{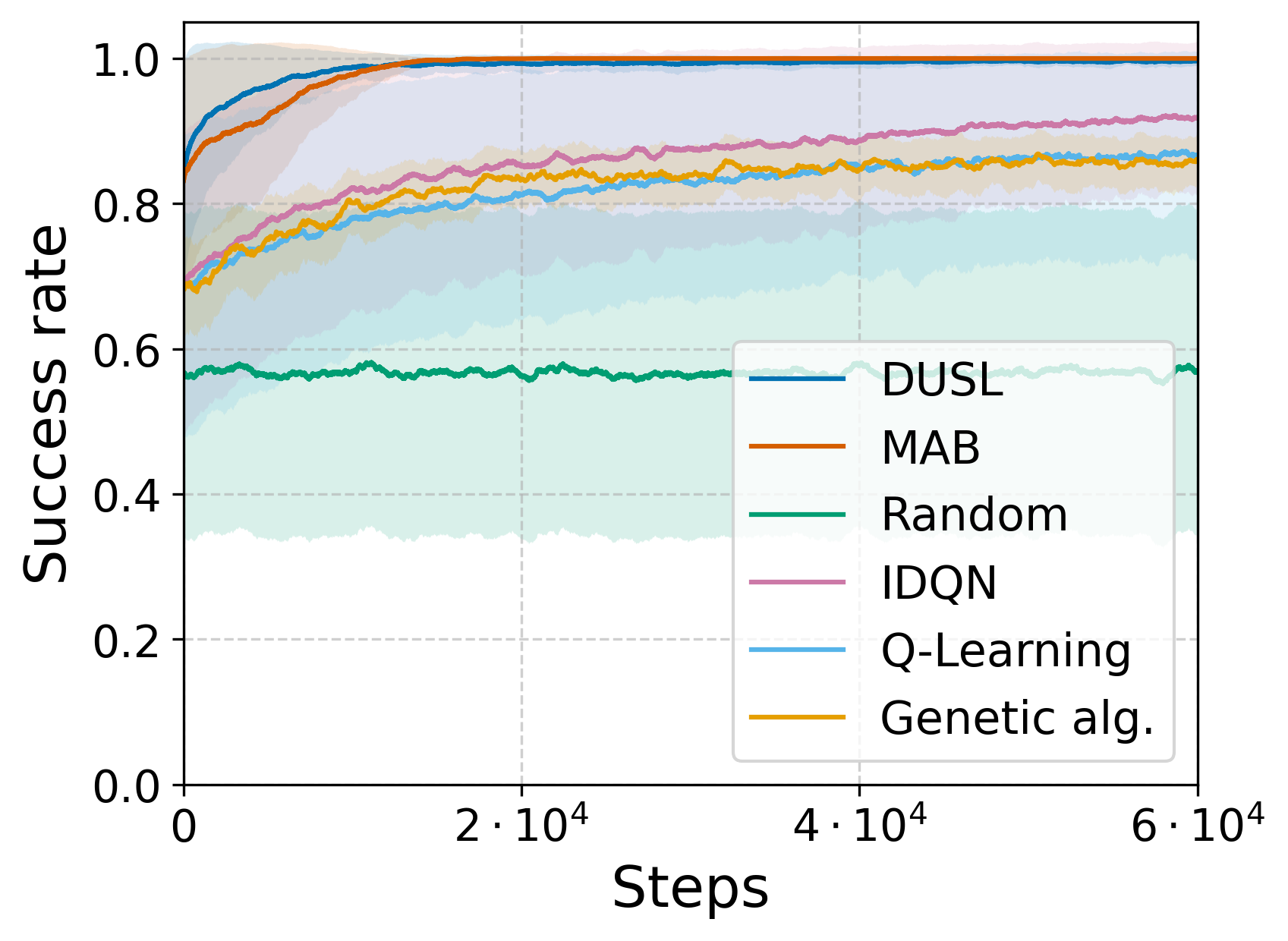}
        \caption{$N = 16$, $M = 4$,\\ $|\mathcal{A}_l| = 4$.}
        \label{fig:cond_off_1_msg_s16}
    \end{subfigure}
    \hfill
    \begin{subfigure}[b]{0.24\linewidth}
        \includegraphics[width=\textwidth]{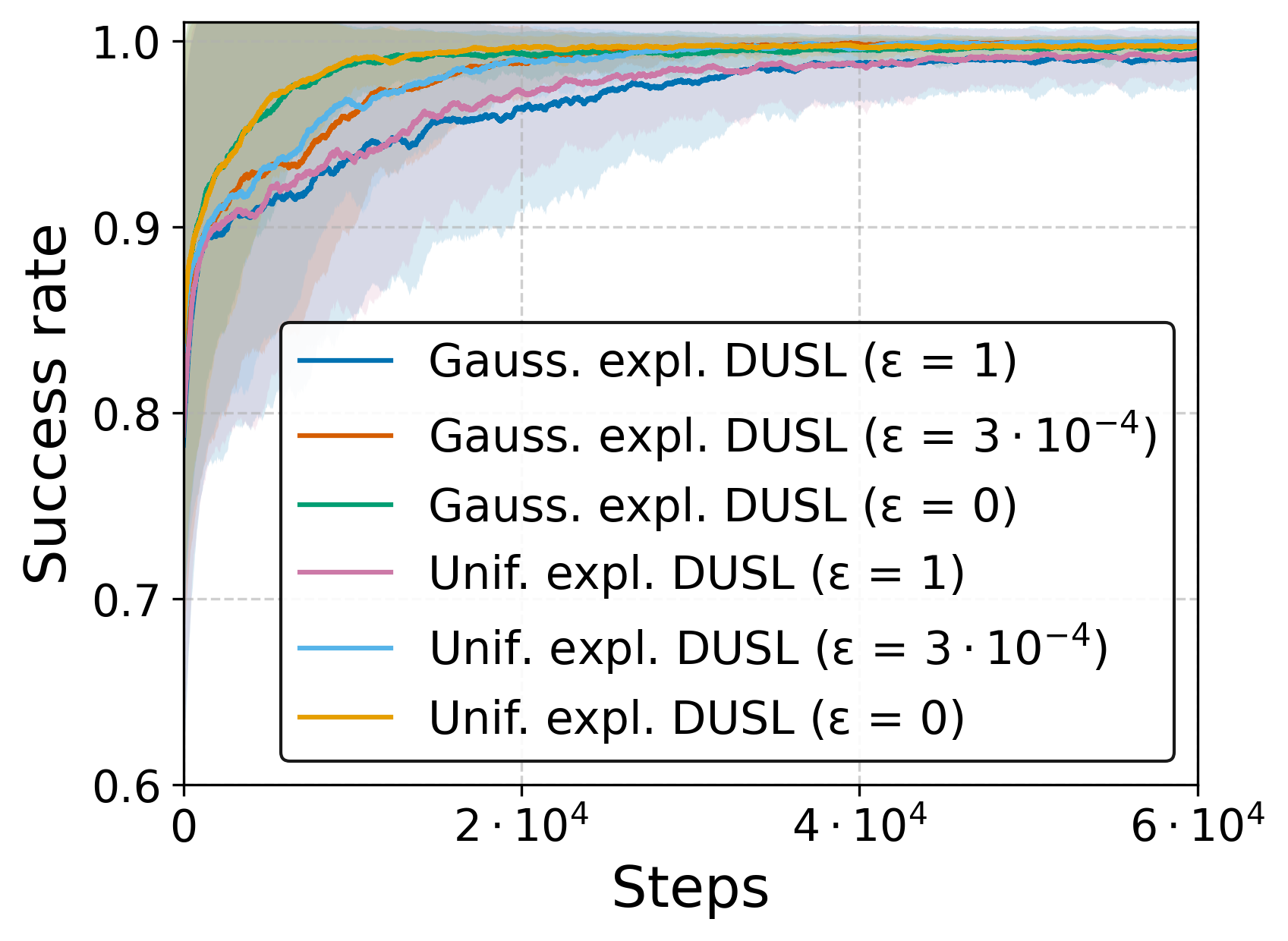}
        \caption{$N = 16$, $M = 4$,\\ $|\mathcal{A}_l| = 4$.}
        \label{fig:cond_off_1_msg_s16_ABL}
    \end{subfigure}
    \hfill
    \begin{subfigure}[b]{0.24\linewidth}
        \includegraphics[width=\textwidth]{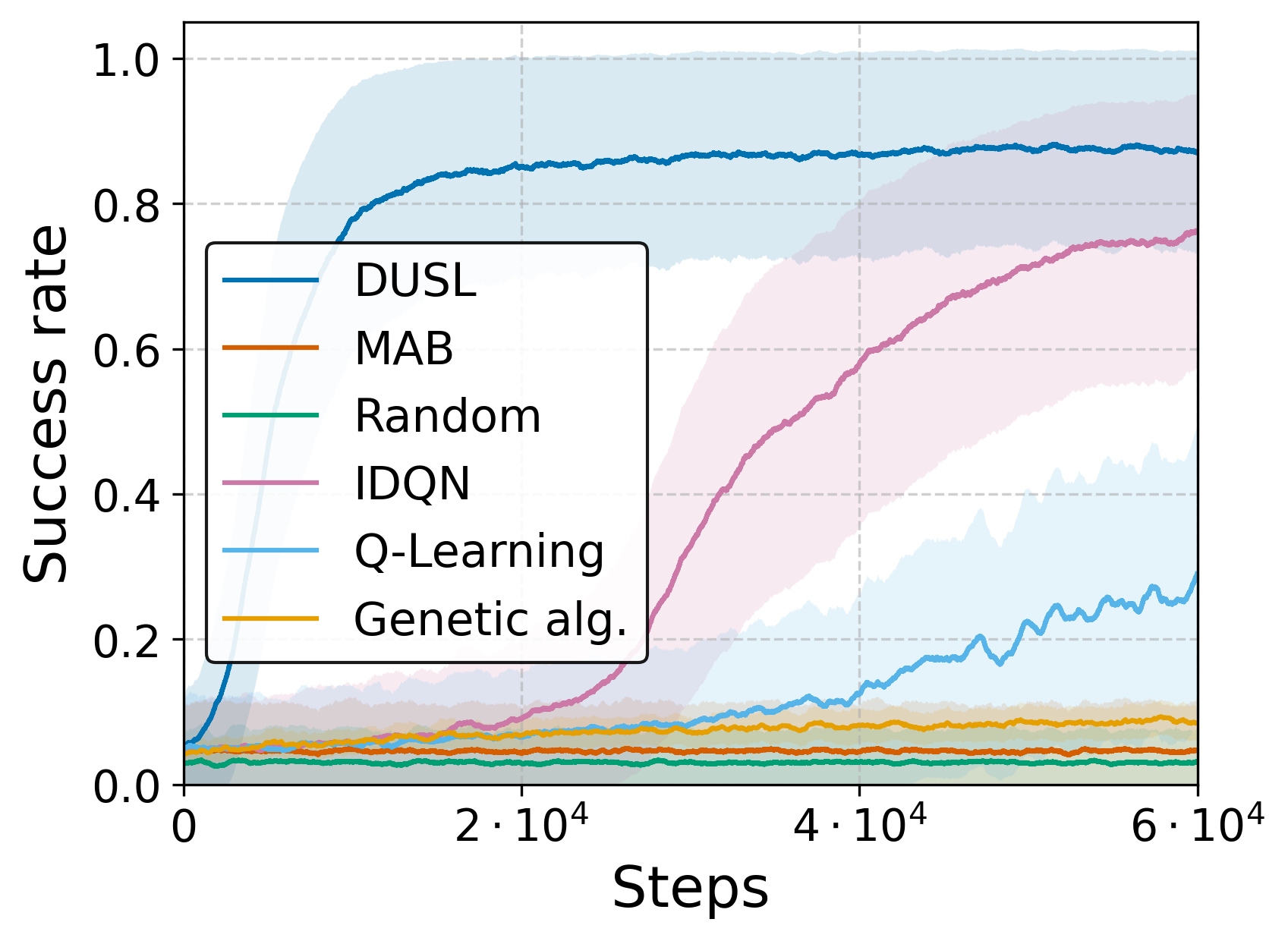}
        \caption{$N = 64$, $M = 16$,\\ $|\mathcal{A}_l| = 12$.}
        \label{fig:cond_off_1_msg_s64}
    \end{subfigure}
    \hfill
    \begin{subfigure}[b]{0.24\linewidth}
        \includegraphics[width=\textwidth]{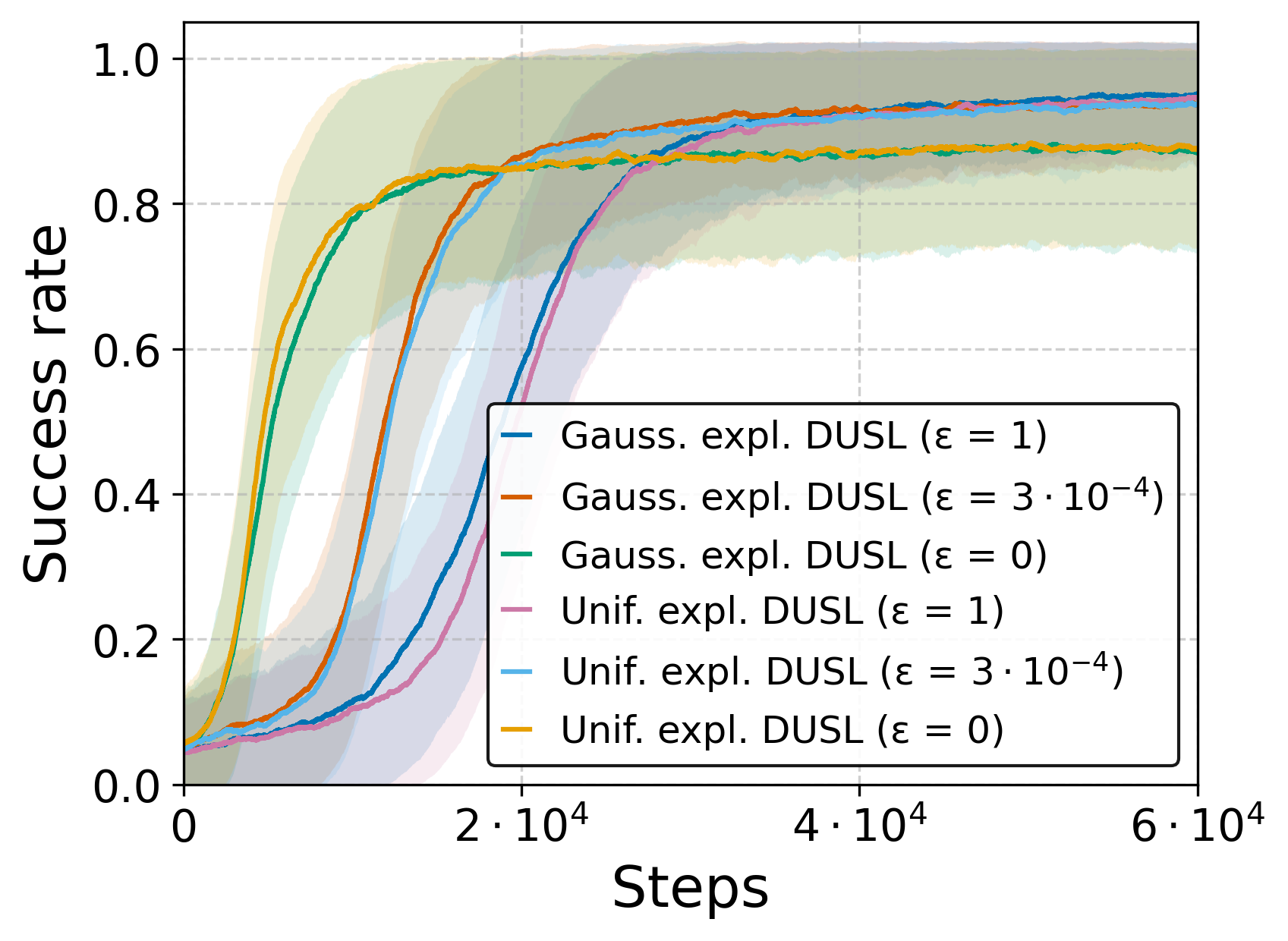}
        \caption{$N = 64$, $M = 16$,\\ $|\mathcal{A}_l| = 12$.}
        \label{fig:cond_off_1_msg_s64_ABL}
    \end{subfigure}
    \hfill
    \caption{\blue{Training in static scenarios with conditional activation patterns and $L = 1$.}}
    \label{fig:cond_off_1_msg}
\end{figure*}

\begin{enumerate}

    \item \textit{Conditional activation}: In this setting, we characterize the activation of a node conditionally to the others. We consider $L$ \acp{BN} to determine the active node set $\mathcal{A}_l$ of message $l \in \mathcal{L}$ for $l=1,\ldots,L$. Each \ac{BN} is modeled by a directed acyclic graph comprising a set of nodes $\mathcal{V} = \{V_l^1, \ldots, V_l^{N_l}\}$ representing $N_l$ variables. 
    These variables are arranged in a linear sequence, where each has at most one incoming and one outgoing edge, forming a path from a source to a sink. Specifically, the variable $V_l^i$ in the sequence determines the $i$-th active node in $\mathcal{A}_l$ and the conditional dependencies between node activations are modeled by conditional probability tables. We get $N_l$ active nodes transmitting message $l$ by sampling from the joint probability function defined by the \ac{BN}, which is given by the chain rule of probability as
    \begin{align}\label{eq:CA}
    \begin{split}
        P(V_l^1, V_l^2, \ldots, V_l^{N_l}) \!=\! \prod_{i=1}^{N_l}\! P(V_l^i \mid V_l^{i-1},..., V_l^1).
    \end{split}
    \end{align}

    \item \textit{General activation}: In this setting, each node in the active node set $\mathcal{A}_l$ of each message $l$ is selected independently from the others. We consider $L$ multinomial distributions to determine $\mathcal{A}_l$ for $l=1,\ldots,L$. Specifically, we sample $\mathcal{A}_l$ through $N_l$ independent draws:
    \begin{align}
    \begin{split}
        \mathcal{A}_l \sim \text{Multinomial}(N_l, d_l^1, \dots, d_l^N),
    \end{split}
    \end{align}
    where $d_l^i$ is the probability that node $i$ is selected in the active set $\mathcal{A}_l$. 
    Consequently, we can bound the cardinality of the active set transmitting $l$ as $\vert \mathcal{A}_l \vert \leq N_l$. 

\end{enumerate}
For both cases, we evaluate 
the proposed 
algorithm for both static and dynamic scenarios. The former considers a fixed distribution of activation patterns $p$ over time, while the latter includes time-varying 
shifts in $p$.
Additional experiments investigating  
unbalanced message availability, temporal correlation, and learned policy statistics are provided in Appendix~\ref{appendix:B}.%}

\subsection{Implementation Details and Baselines}

Each DNN used for testing DUSL consists of three layers with layer dimensions [256, 128, $M$]. The model is trained using the Adam optimizer with a learning rate $\lambda= 10^{-3}$, and employs ReLU activations. 
\blue{We designed these shallow neural models to prioritize computational efficiency and practical deployability. In our tests on standard multi-threading computing units (i.e., Raspberry Pi 4), the execution times for the inference steps of DUSL consistently accounted for less than a millisecond.} 
We also investigate a variant of our DUSL algorithm, where each node deploys a single \ac{DNN} that generates $L\,M$ output neurons to simultaneously determine its transmission choices for all messages. 
{Training stability within the policy-gradient procedure is maintained through several mechanisms. First, the binary reward signal ($\xi \in \{0, 1\}$) naturally bounds gradient updates. 
Second, leveraging the Adam optimizer, which features adaptive and diminishing step sizes, balances convergence speed and stability. 
Third, the exploration decay factor $\varepsilon$ controls the variance of the input, ensuring a smooth transition from exploration to exploitation. Prior to training, the DUSL framework initialize hidden layers by Kaiming uniform initialization. Crucially, this centers the pre-activation outputs near zero, causing the sigmoid function to yield initial transmission probabilities near $0.5$. This setup guarantees initial exploration across all communication opportunities and prevents the policy-gradient algorithm from stalling due to saturated output neurons.} 

For baselines, %the most relevant one  
% is
\blue{we compare our approach against a comprehensive set of five baselines: %alternative methods: 
(i) }
the distributed \ac{MAB} approach leveraging Thompson sampling proposed in \cite{learningToSpeak}, where each active node chooses which opportunities to use for message transmission in a
distributed manner. In this approach, the action space consists of all possible \(2^M\) subsets of communication opportunities, and each node learns a policy over this combinatorial space through bandit feedback, resulting in a higher learning complexity as discussed in Sec.~\ref{subsec:complexity}\blue{; (ii) a \ac{GA} \cite{gaRef}; (iii) a tabular Q-Learning approach \cite{qlearnRef}; (iv) an \ac{IDQN} method \cite{idqnRef}; and (v)} 
a random strategy, where each node selects a random subset of opportunities for transmission of each message $l \in \mathcal{L}$ at each time step, with each opportunity being independently chosen with probability \( \frac{1}{|\mathcal{A}|} \).
While this na\"ive approach does not leverage any learning or coordination, it serves as a useful lower-bound reference. 
Across multiple runs, fixed random seeds are used to ensure reproducibility.

\subsection{Conditional Activation Patterns}\label{subsec:CA}

In this section, we first evaluate our method 
in static scenarios for different numbers of 
nodes $N$, channels $M$, and messages $L$ under conditional activation patterns, and then consider dynamic scenarios in which the distribution of active node sets $p$ changes over time, resulting in different conditional activation patterns. In each of the following cases, we average performance curves over 15 different independent runs to ensure the robustness of the methodology. 

\textit{\textbf{Scenario definition.}} To construct the \ac{BN} for each message $l \in \mathcal{L}$, a hierarchical sampling process defines the active node set $\mathcal{A}_l$. We begin with four random root nodes, each activating up to four child nodes, continuing recursively to form a tree representing conditional dependencies. 
To maintain computational tractability and avoid an excessively large set of active nodes, we introduce constraints on the tree's expansion at deeper levels. Specifically:
(i) Up to level 4, each node can activate four child nodes;
(ii) From level 5 to level 8, each node activates at most three child nodes;
(iii) Beyond level 8, the branching is limited to two child nodes per node.
This controlled growth ensures that the resulting \ac{BN} remains compact and the sampled active sets \( \mathcal{A}_l \) are of manageable size while preserving a rich structure of conditional dependencies.

\textit{\textbf{Static scenarios.}} Fig.~\ref{fig:cond_off_1_msg} presents the performance of our algorithm for a single message, i.e., \( L = 1 \), under two different setups. The first one in Fig.~\ref{fig:cond_off_1_msg_s16} comprises a small scenario with $N = 16$ and $M = 4$, where the size of active node set is $|\mathcal{A}_1| = 4$. 
\blue{{We observe that the proposed DUSL exhibits the fastest convergence speed, while MAB converges more slowly although eventually reaching convergence. 
The other baselines (i.e., GA, tabular Q-Learning, and IDQN) fall noticeably behind with slower convergence speeds even in this simple scenario.}} 
\blue{As we increase the number of nodes to $64$, the scenario becomes more challenging, }%However, 
%as shown in Fig.~\ref{fig:cond_off_1_msg_s64}, %if we increase the number of nodes to $64$, 
\blue{Fig.~\ref{fig:cond_off_1_msg_s64} shows that }the \ac{MAB} \blue{as well as GA and tabular Q-Learning} 
do not scale and 
\blue{fall significantly behind or completely fail to solve the optimization problem}. 
\blue{{In contrast},  
the deep models (DUSL and IDQN) exhibit good performance, with DUSL consistently staying above IDQN in terms of overall performance.} 
In Figs.~\ref{fig:cond_off_1_msg_s16_ABL} and \ref{fig:cond_off_1_msg_s64_ABL}, it is possible to evaluate the impact of the exploration decay factor $\varepsilon$: if $\varepsilon = 1$, i.e., almost no exploration noise is fed to network, the learning process is slower, but in the end, achieves better performance than 
the case with 
$\varepsilon = 0$. In the latter case, 
the input fed to the neural network is always noisy, facilitating exploration at early training phases, but also preventing to converge to better values at the end of the training. A good trade off is to set a decaying $\varepsilon$, such that the training 
is faster at the beginning and the final convergence is not compromised. 
\blue{Furthermore, we performed ablation studies on the sampling distribution of the exploration parameter $s$, comparing sampling $s$ from a Gaussian random variable or from an uniform random variable. We verified that for conditional patterns, changing the distribution of $s$ does not remarkably impact the overall convergence performance.}

\begin{figure}[t]
    \centering
    \begin{subfigure}[b]{0.48\columnwidth}
        \centering
        \captionsetup{justification=centering}
        \includegraphics[width=\textwidth]{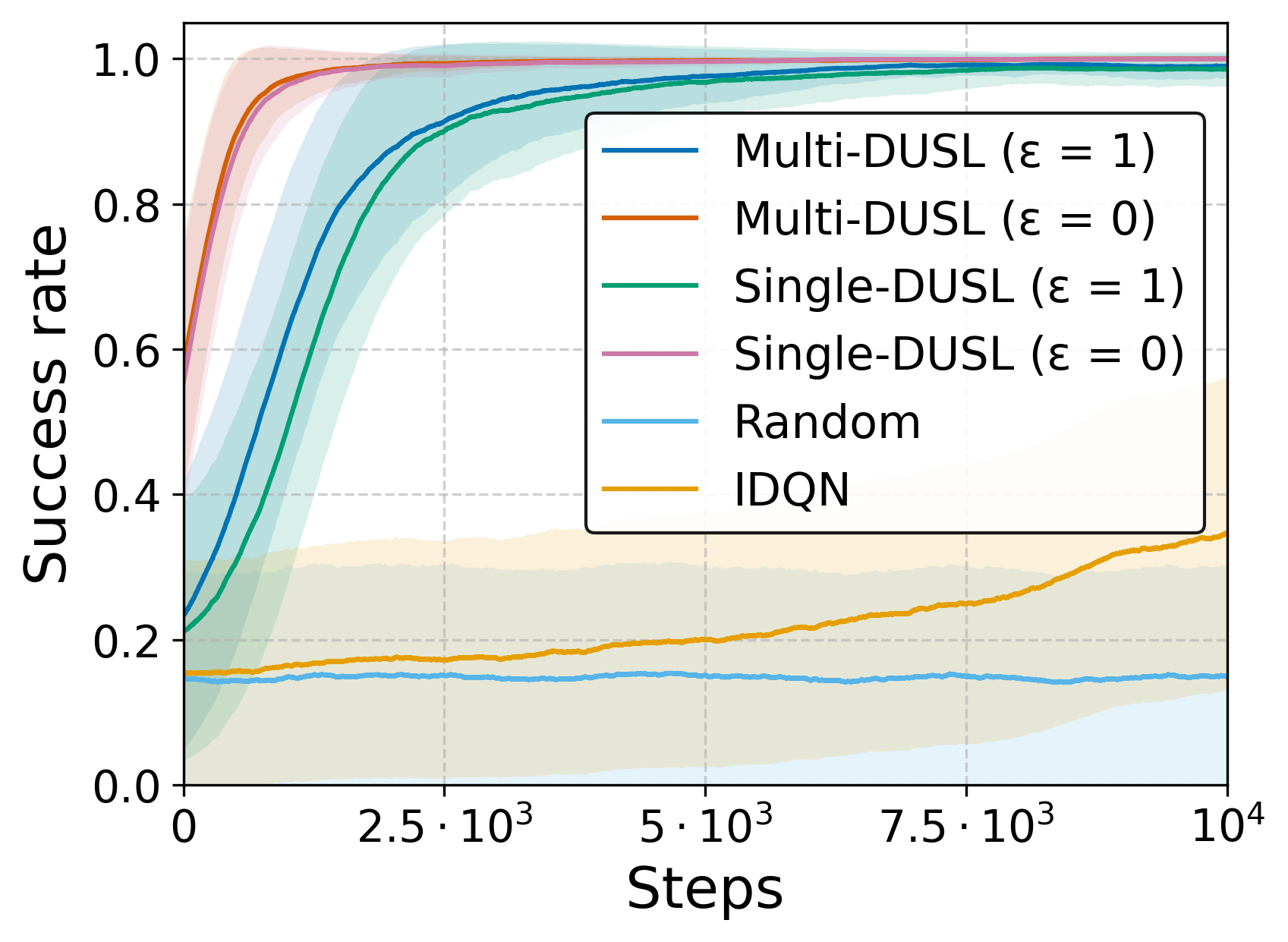}
        \caption{$N = 64$, $M = 32$, \\ $|\mathcal{A}_l| = 4$, $L = 2$.}
        \label{fig:cond_off_2_msg}
    \end{subfigure}
    \hfill
    \begin{subfigure}[b]{0.48\columnwidth}
    \captionsetup{justification=centering}
        \includegraphics[width=\textwidth]{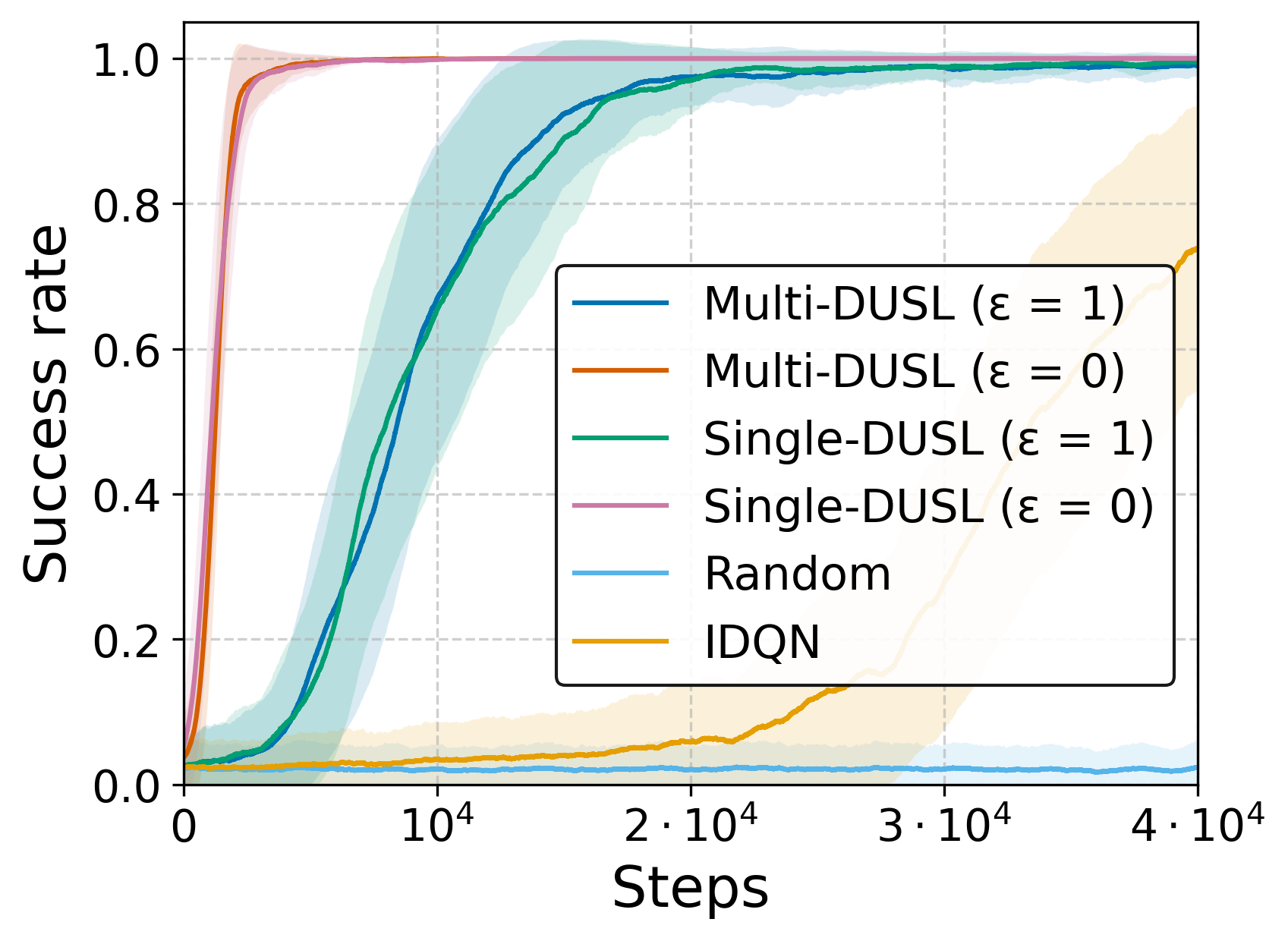}
        \caption{$N = 64$, $M = 64$, \\ $|\mathcal{A}_l| = 2$, $L = 4$.}
        \label{fig:cond_off_4_msg}
    \end{subfigure}
    \caption{\blue{Training in static scenarios with conditional activation patterns and multiple messages.}}
    \label{fig:cond_off_2-4_msg}
\end{figure}

In Fig.~\ref{fig:cond_off_2-4_msg}, we extend the scenario to multiple messages (\( L = 2 \) and \( L = 4 \)). The configurations of the first case in Fig. \ref{fig:cond_off_2_msg} include \( N = 64 \) nodes and \( M = 32 \) opportunities with \( |A_l| = 4 \) active nodes per message, and that of the second case in Fig. \ref{fig:cond_off_2-4_msg} include \( N = 64 \) nodes and \( M = 64 \) opportunities with \( |A_l| = 2 \). With multiple messages, the complexity of the problem increases as the nodes must coordinate to successfully send all the messages. 
In these large-scale settings, we omit to present the distributed \ac{MAB} \blue{and simpler tabular/evolutionary approaches} as \blue{they are} already proved to be ineffective for $L=1$ and $N=64$. We note that \ac{MAB} becomes more computational and memory inefficient for greater $M$. 
For instance, with \( M = 32 \), each node would need to consider \( 2^{32} \) possible arms, requiring at least \( 2^{32} \) steps to sample each arm even once, and maintaining at least \( 2 \cdot 2^{32} \) counters to track successes and failures, which is clearly impractical in terms of both convergence time and memory requirements. 
\blue{The deep IDQN baseline was the most robust among the alternative methods, yet it still features significantly worse performance than {our} DUSL across both $L=2$ and $L=4$ configurations.}

We also investigate a DUSL variant where each node deploys a single DNN with $L M$ outputs for the joint assignment of all $L$ messages to $M$ resources. Fig.~\ref{fig:cond_off_2-4_msg} compares \emph{Single-DUSL} (one DNN per node) against \emph{Multi-DUSL} (the original design with $L$ independent DNNs per node). 
Empirically, 
\blue{Single-DUSL and Multi-DUSL deliver comparable final success rates, though Single-DUSL exhibits slightly slower convergence in certain configurations.} 
We attribute this behaviour to the stronger coupling induced by joint assignment: 
a single network must learn to discriminate among \(L\,M\) decision targets and its neurons jointly contribute to all messages, which increases the learning complexity 
compared to having one network dedicated to each message. Although the single-DNN design reduces the number of distinct parameter sets, in our large-scale settings the reduced per-message expressivity and the more challenging optimization outweighed those advantages. 
For the remainder of the experiments we focus on Multi-DUSL 
and, for brevity, simply refer to it as \emph{DUSL}.

\begin{figure}[t]
    \centering
    \begin{subfigure}[b]{0.48\columnwidth}
        \captionsetup{justification=centering}
        \includegraphics[width=\textwidth]{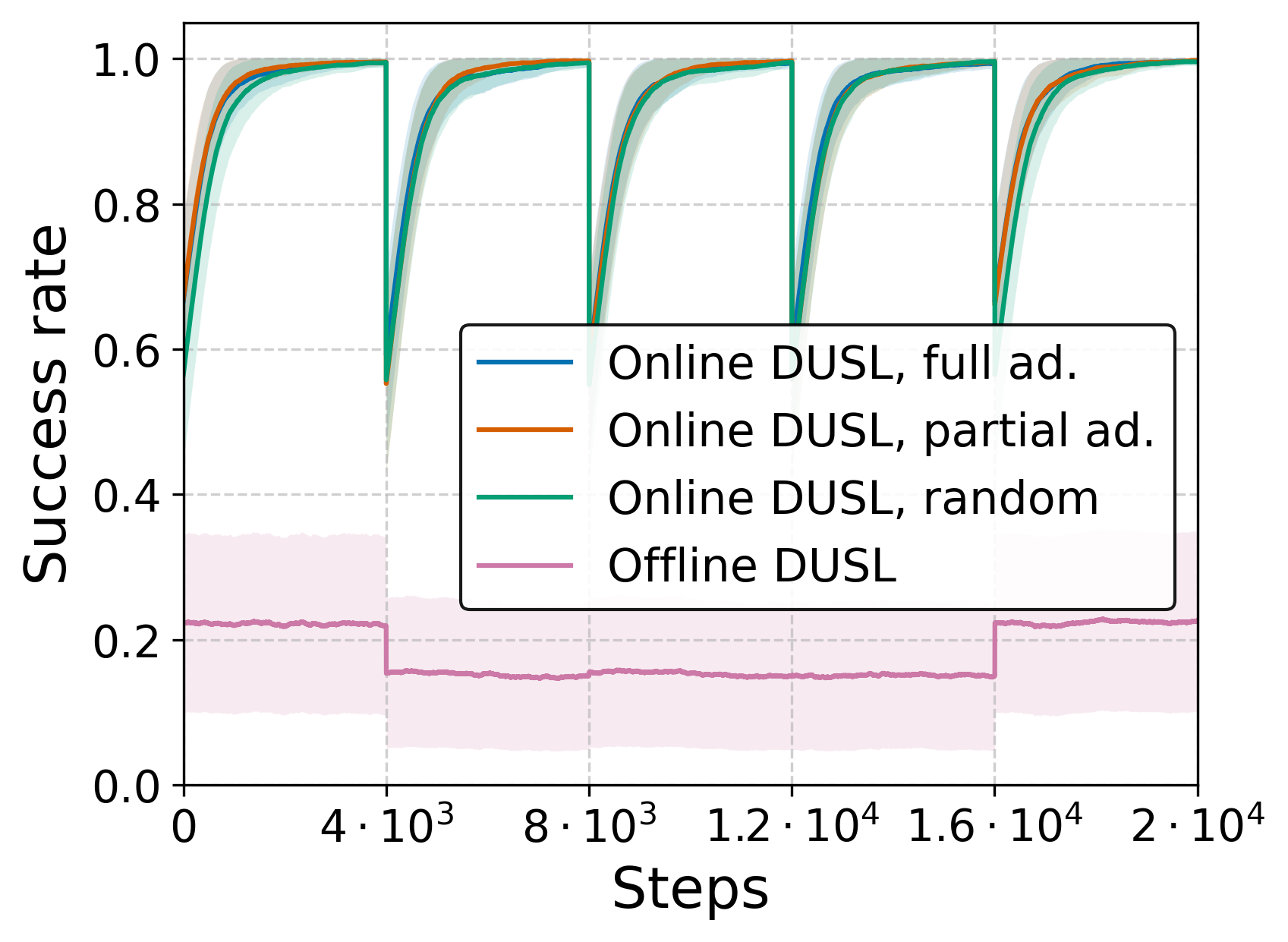}
        \caption{$N = 64$, $M = 32$, \\ $|\mathcal{A}_l| = 4$, $L = 2$.}
        \label{fig:cond_fft_2_msg}
    \end{subfigure}
    \hfill
    \begin{subfigure}[b]{0.48\columnwidth}
        \captionsetup{justification=centering}
        \includegraphics[width=\textwidth]{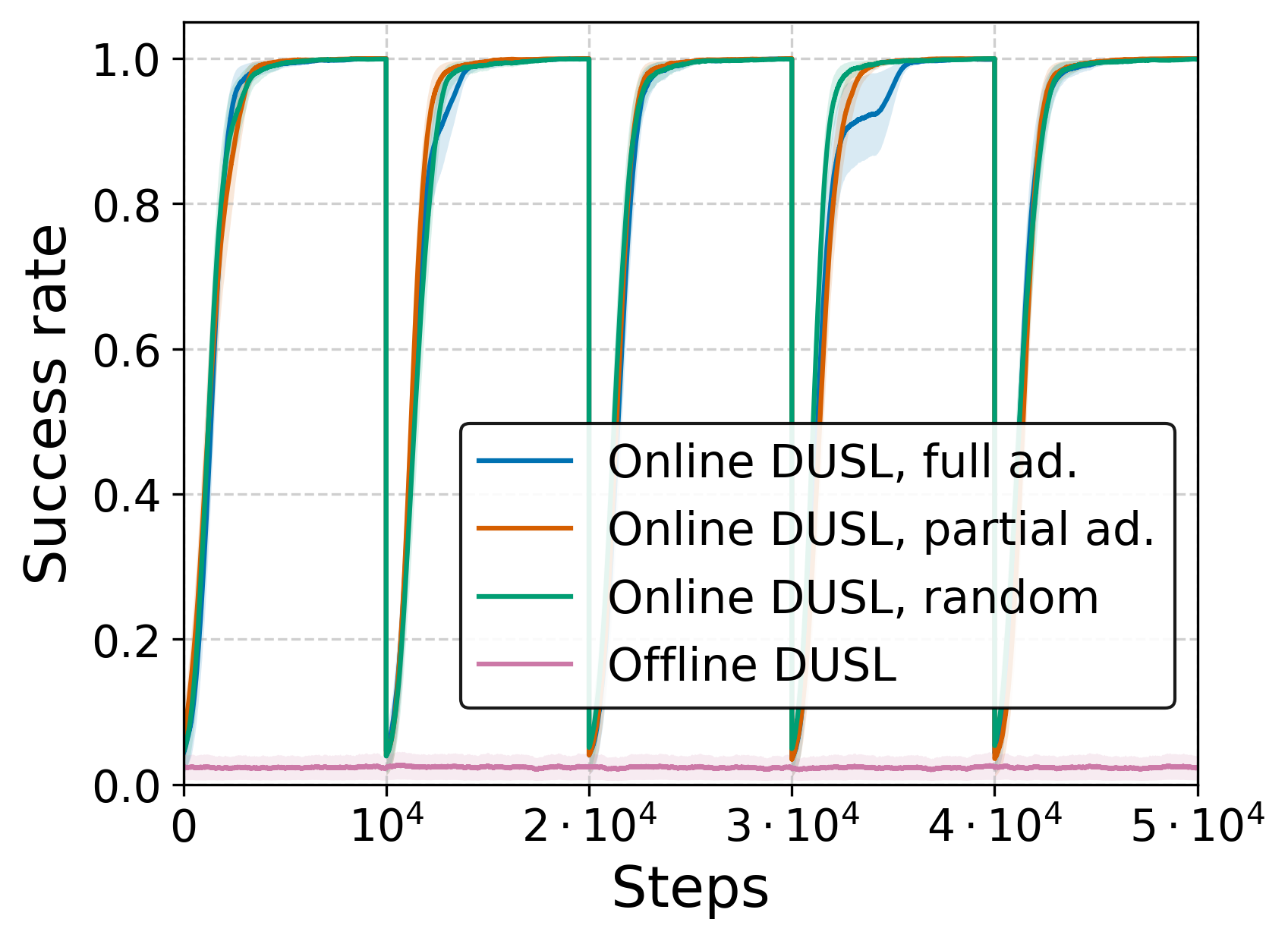}
        \caption{$N = 64$, $M = 64$, \\ $|\mathcal{A}_l| = 2$, $L = 4$.}
        \label{fig:cond_fft_4_msg_fft}
    \end{subfigure}
    \caption{{Adaptation to dynamic scenarios with conditional activation patterns and multiple messages, where 5 changes of $p_A$ occur.
    }}
    \label{fig:cond_fft_2-4_msg}
\end{figure}

\textit{\textbf{Dynamic scenarios.}} In Fig.~\ref{fig:cond_fft_2-4_msg}, we examine adaptation capabilities of the proposed 
online learning mechanism in dynamic scenarios, where nodes are required to quickly adjust their transmission strategies according to variations in the activation distribution $p$. 
We set $N$, $M$, and $L$ as that in 
Fig.~\ref{fig:cond_off_2-4_msg}, 
and change the environment periodically, i.e., allow the model to adjust 
within a very limited amount of time after the environment changes. We test the following four variants: 
i) DUSL with full adaptation, where all parameters of the offline trained 
model 
are updated online; 
ii) DUSL with partial adaptation, where the first hidden layer of the offline trained model is frozen and the rest parameters are updated online; 
iii) DUSL without offline training, where all parameters are randomly initialized and updated online; iv) Offline trained DUSL without online learning. 

We see that the proposed online learning mechanism adapts the offline trained policy quickly to distribution shifts and maintains the performance of DUSL in dynamic scenarios. Online DUSL initialized at the offline trained model achieves the best performance, compared to the other variants, where partial adaption reduces the learning complexity and improves the adaption rate because it updates only a subset of model parameters online. 
% Moreover, 
We remark that the changes in $p$ are harsh and each change is highly uncorrelated 
to the other in this scenario 
[cf. \eqref{eq:CA}], as the sampling form the \ac{BN} is uniform and there is a limited number of possible active sets. This 
highlights the robustness of the online learning mechanism in challenging dynamic scenarios, 
which leads to significant performance improvements for each change of $p$.

\subsection{General Activation Patterns}

In this section, we evaluate the proposed method in general activation patterns, 
where there are no significant dependencies between the nodes' activations. 
We consider both static 
and dynamic scenarios, and average performance over 15 different independent runs to ensure robustness. 

\textit{\textbf{Scenario definition.}} As discussed, we use $L$ multinomial distributions to model the activation patterns for messages 
$\mathcal{L}$. For each $l$, the probabilities of the multinomial distribution \( (d_l^1, \dots, d_l^N) \) are sampled from a Dirichlet distribution:
\begin{equation}
	(d_l^1, \dots, d_l^N) \sim \text{Dir}(\mathbf{1}_N)~,
\end{equation}
where \( \mathbf{1}_N \) is a vector of ones of length \( N \). This is a common case for a Dirichlet distribution, where all of the concentration parameters have the same value, resulting in no prior knowledge favoring one node activation over another. Particularly, when all concentration parameters are equal to one, the Dirichlet distribution is equivalent to a uniform distribution over the open standard ($K - 1$)-simplex, i.e., it is uniform over all points in its support.

\begin{figure*}[t]
    \centering
    \begin{subfigure}[b]{0.24\linewidth}
        \includegraphics[width=\textwidth]{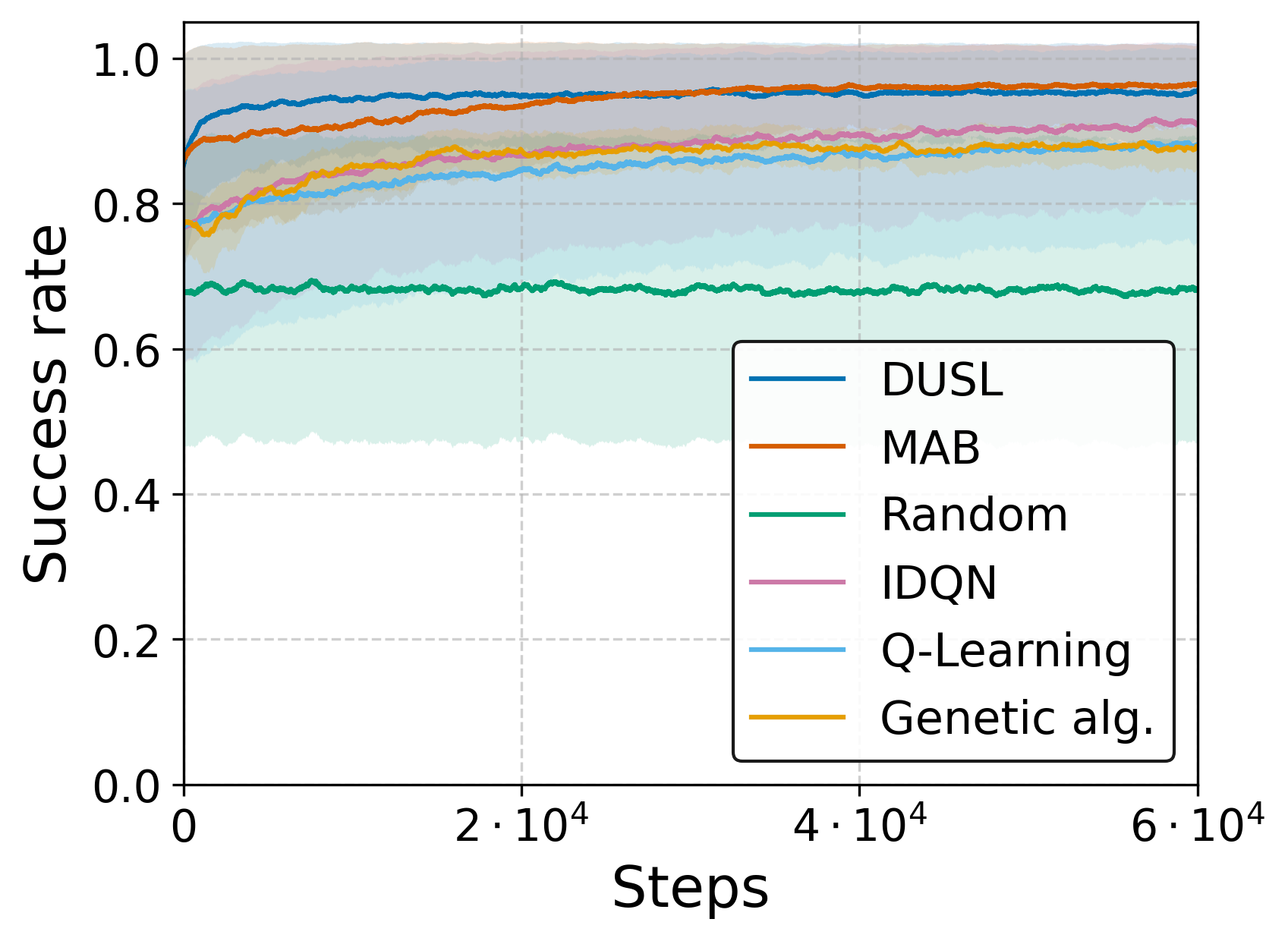}
        \caption{$N = 16$, $M = 4$,\\ $|\mathcal{A}_l| = 4$.}
        \label{fig:general_off_1_msg_s16}
    \end{subfigure}
    \hfill
    \begin{subfigure}[b]{0.24\linewidth}
        \includegraphics[width=\textwidth]{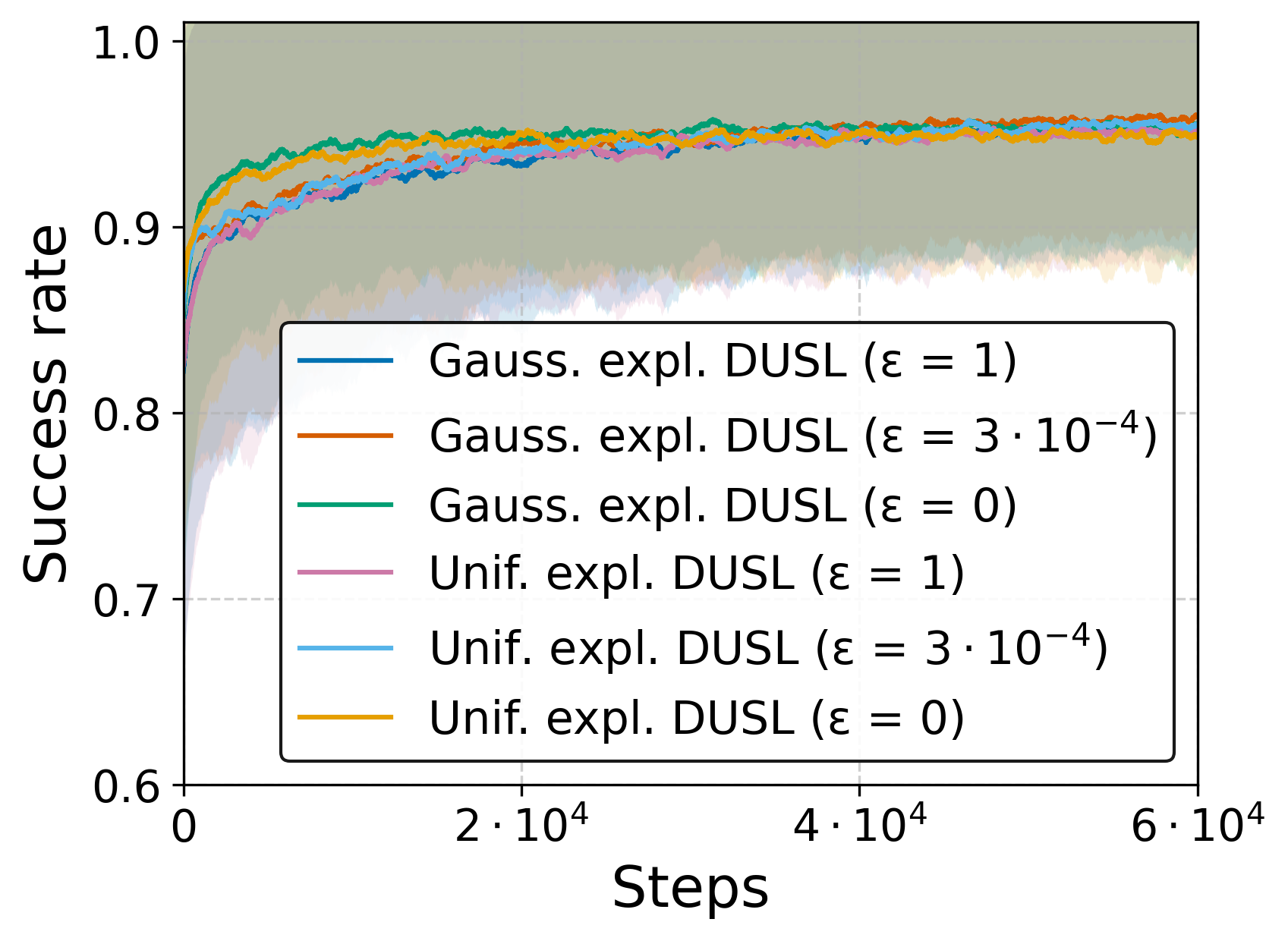}
        \caption{$N = 16$, $M = 4$,\\ $|\mathcal{A}_l| = 4$.}
        \label{fig:general_off_1_msg_s16_ABL}
    \end{subfigure}
    \hfill
    \begin{subfigure}[b]{0.24\linewidth}
        \includegraphics[width=\textwidth]{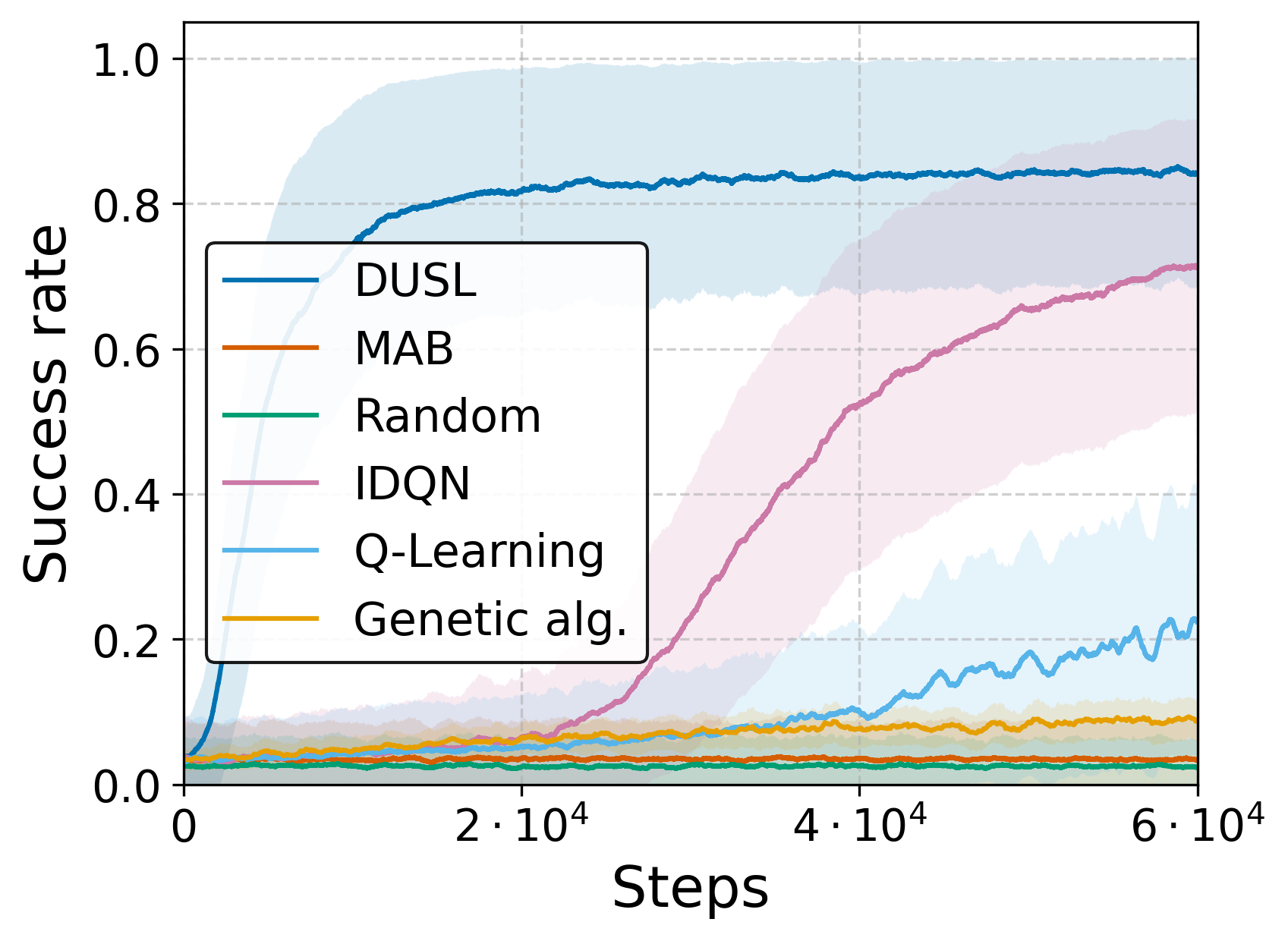}
        \caption{$N = 64$, $M = 16$,\\ $|\mathcal{A}_l| = 16$.}
        \label{fig:general_off_1_msg_s64}
    \end{subfigure}
    \hfill
    \begin{subfigure}[b]{0.24\linewidth}
        \includegraphics[width=\textwidth]{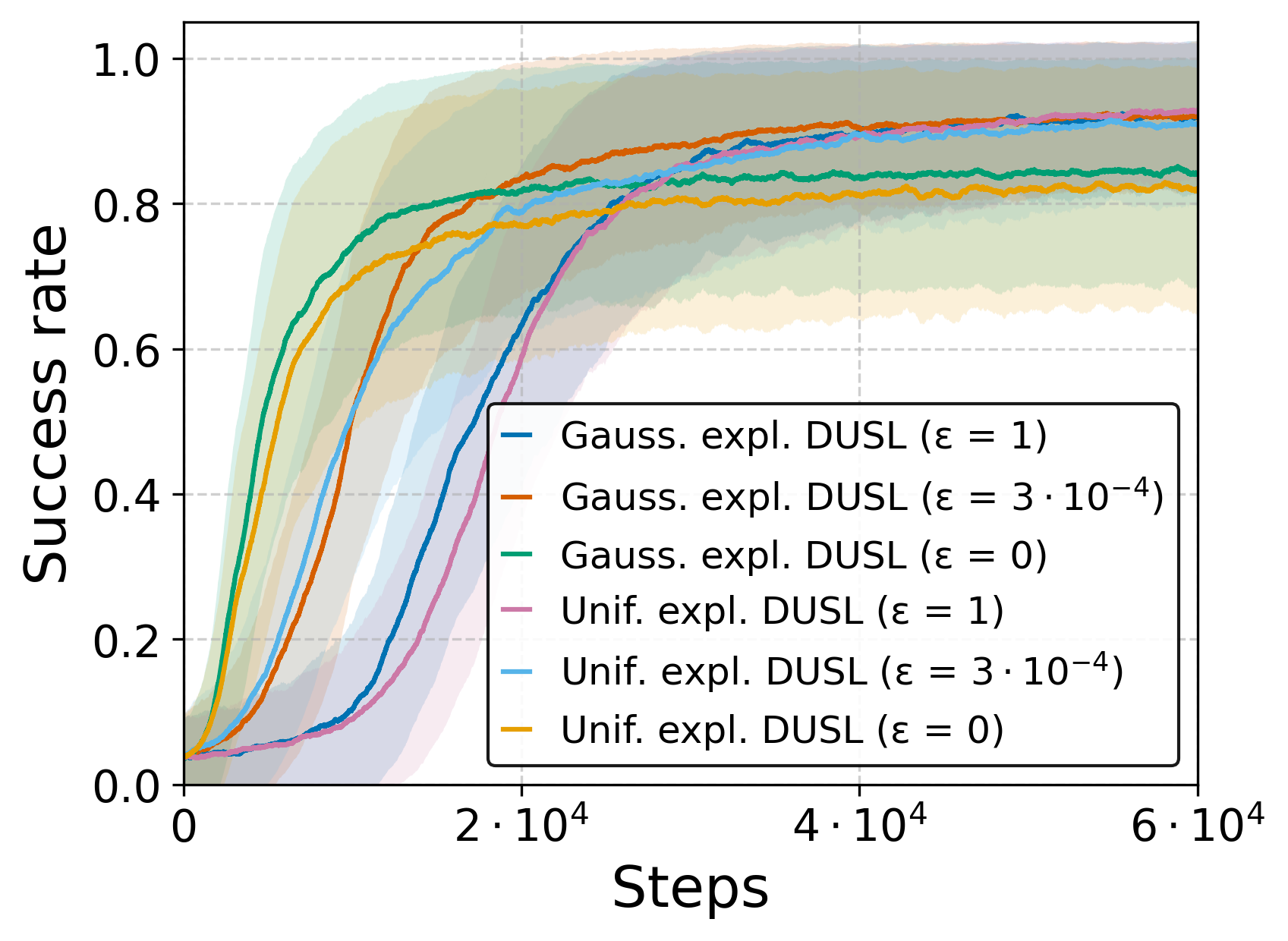}
        \caption{$N = 64$, $M = 16$,\\ $|\mathcal{A}_l| = 16$.}
        \label{fig:general_off_1_msg_s64_ABL}
    \end{subfigure}
    \caption{\blue{Training in static scenarios with general activation patterns and $L = 1$.}}
    \label{fig:general_off_1_msg}
\end{figure*}

\textit{\textbf{Static scenarios.}} In Fig.~\ref{fig:general_off_1_msg}, we show the performance of DUSL in general activation setting for \( L = 1 \) for static scenarios. 
\blue{Similar to the conditional pattern case, the simpler scenario in Fig.~\ref{fig:general_off_1_msg_s16} can be solved by DUSL and MAB, which achieve improved performance and convergence speed than \ac{GA}, tabular Q-Learning, and \ac{IDQN}.} 
However, in the more complex scenario shown in Fig.~\ref{fig:general_off_1_msg_s64} where we increase the number of nodes from 16 to 64, DUSL 
\blue{consistently outperforms IDQN, while the remaining approaches fail to scale}.
In Figs.~\ref{fig:general_off_1_msg_s16_ABL} and \ref{fig:general_off_1_msg_s64_ABL} the exploration decay factor \( \varepsilon \) has a notable impact; 
with reduced exploration, i.e., $\varepsilon = 1$, the training is slower but in the end achieves better performance than the case in which $\varepsilon = 0$. 
\blue{Furthermore, these figures show a {slight} 
improvement in sampling $s$ from a Gaussian distribution rather than a uniform distribution in this more complex scenario.}

\begin{figure}[t]
    \centering
    \begin{subfigure}[b]{0.48\columnwidth}
        \centering
        \captionsetup{justification=centering}
        \includegraphics[width=\textwidth]{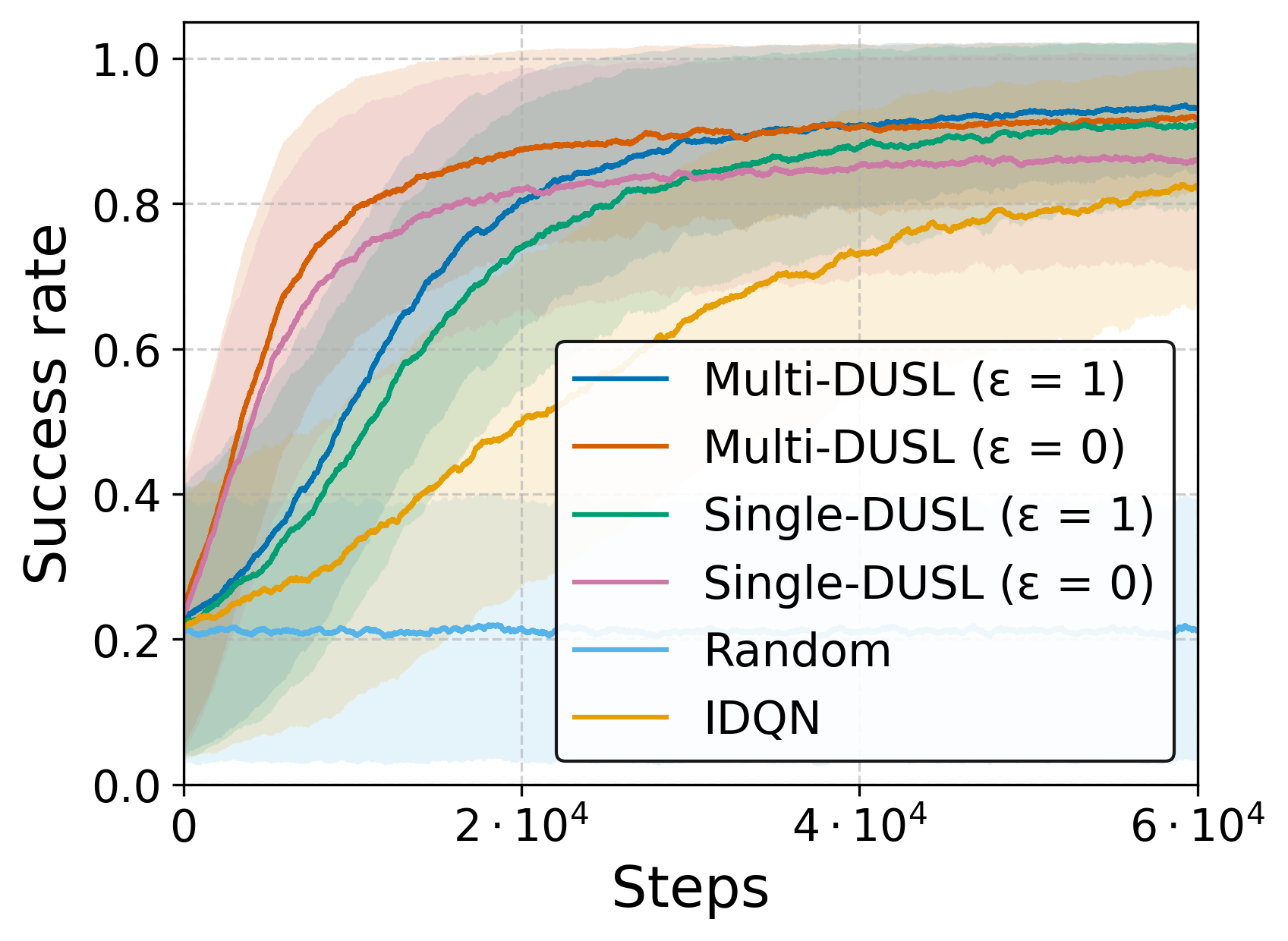}
        \caption{$N = 64$, $M = 32$, \\ $|\mathcal{A}_l| = 4$, $L = 2$.}
        \label{fig:general_off_2_msg}
    \end{subfigure}
    \hfill
    \begin{subfigure}[b]{0.48\columnwidth}
        \centering
        \captionsetup{justification=centering}
        \includegraphics[width=\textwidth]{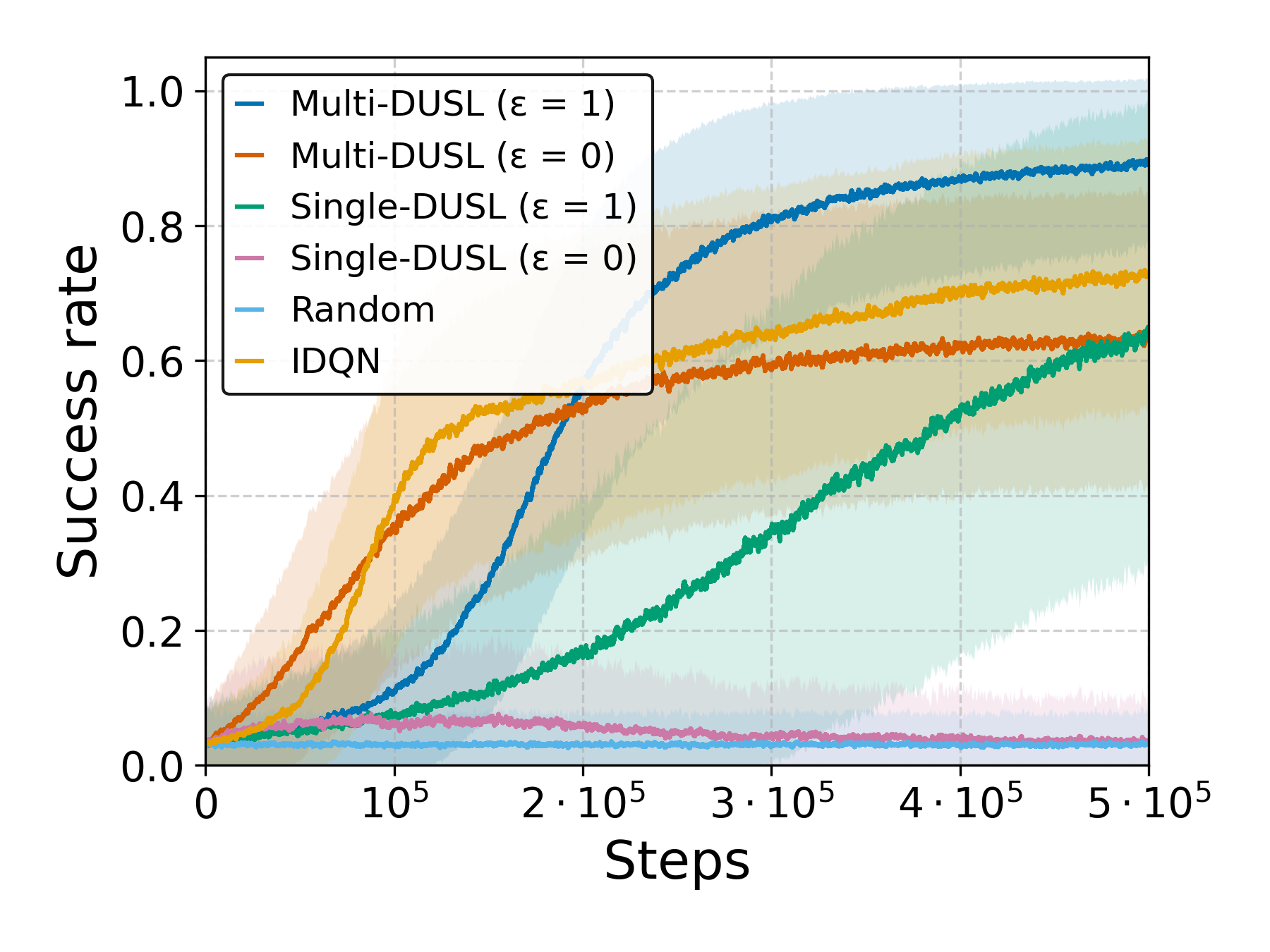}
        \caption{$N = 64$, $M = 64$, \\ $|\mathcal{A}_l| = 2$, $L = 4$.}
        \label{fig:general_off_4_msg}
    \end{subfigure}
    \caption{\blue{Training in static scenarios with general activation patterns and multiple messages.}}
    \label{fig:general_off_2-4_msg}
\end{figure}

In Fig.~\ref{fig:general_off_2-4_msg}, we extend the analysis to multiple messages, considering \( L = 2 \) and \( L = 4 \). The first scenario in Fig.~\ref{fig:general_off_2_msg} features \( N = 64 \), \( M = 32 \), and \( |A_l| = 4 \) active nodes per message, while the second in Fig.~\ref{fig:general_off_4_msg} presents \( N = 64 \), \( M = 64 \), and \( |A_l| = 2 \). Unlike the conditional case, the nodes must coordinate their transmissions without any structured dependencies to rely on. Consequently, the general activation case is more difficult, and convergence is slower. 
In these cases, we omit results for the distributed \ac{MAB} \blue{and simpler tabular/evolutionary approaches} due to similar reasons as those discussed 
in Sec. \ref{subsec:CA}: \blue{they already prove to be ineffective for $L=1$ and $N=64$}, and \ac{MAB} results to be computationally and memory inefficient for greater $M$ too.

\blue{Crucially, in the general activation setting, we observe a significant divergence in performance between Single-DUSL and Multi-DUSL. As shown in Fig.~\ref{fig:general_off_2-4_msg}, Multi-DUSL ($\varepsilon=1$) consistently exhibits superior performance compared to Single-DUSL in terms of both convergence speed and final achieved value. 
This divergence underscores the inherent advantages of task specialization under a general activation pattern. A joint neural architecture like Single-DUSL struggles with the vast action space, requiring significantly more neurons and structural complexity to achieve the same tasks. By contrast, Multi-DUSL leverages specialized models to robustly solve the coordination problem. Importantly, because both proposed solutions rely on intentionally shallow neural designs, they remain highly efficient; execution times for both Single-DUSL and Multi-DUSL consistently account for less than a millisecond on multi-threading computing units (i.e., Raspberry Pi 4).
}

\begin{figure}[t]
    \centering
    \begin{subfigure}[b]{0.48\columnwidth}
        \centering
        \captionsetup{justification=centering}
        \includegraphics[width=\textwidth]{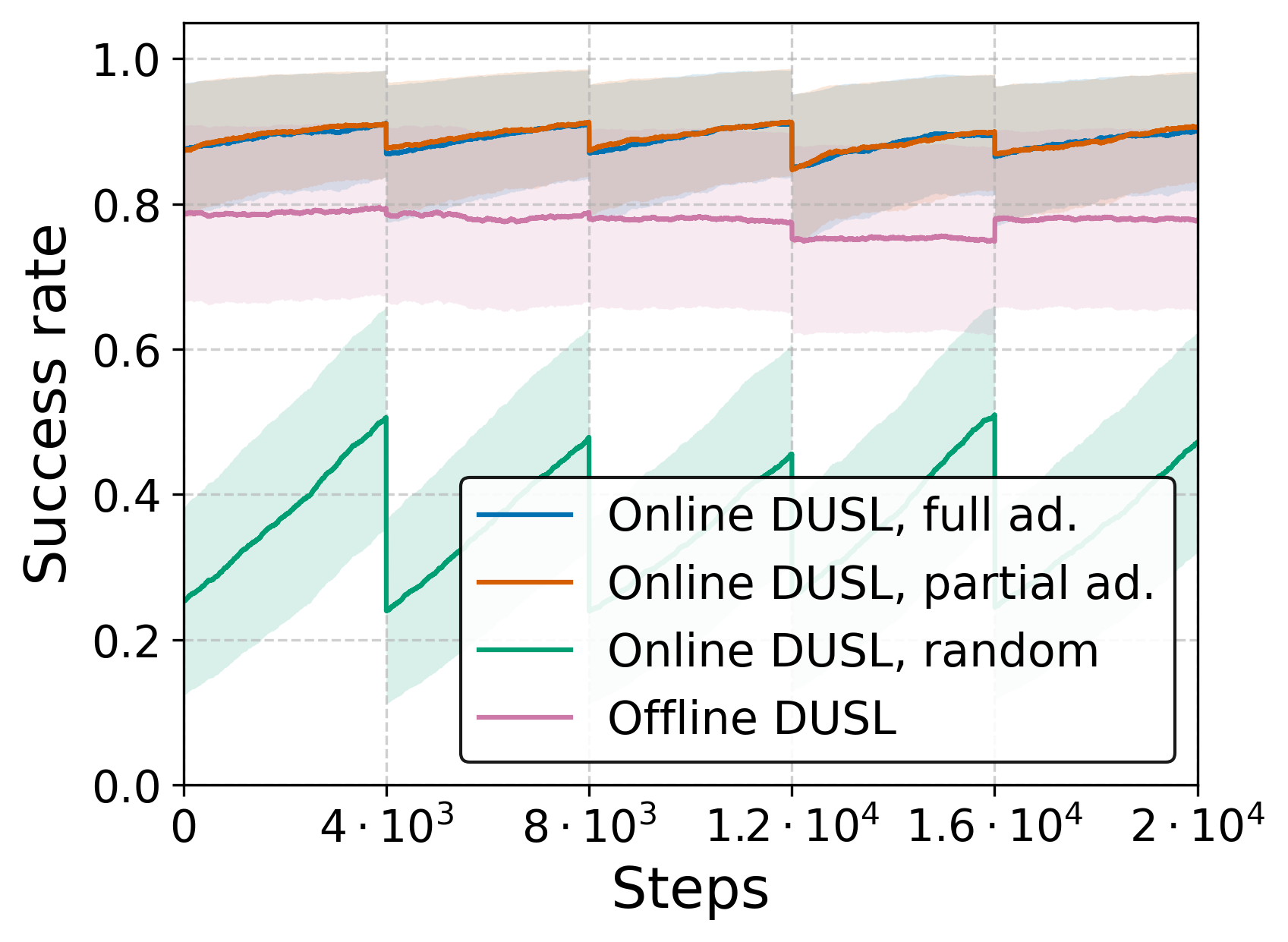}
        \caption{$N = 64$, $M = 32$, \\ $|\mathcal{A}_l| = 4$, $L = 2$.}
        \label{fig:general_fft_2_msg}
    \end{subfigure}
    \hfill
    \begin{subfigure}[b]{0.48\columnwidth}
        \centering
        \captionsetup{justification=centering}
        \includegraphics[width=\textwidth]{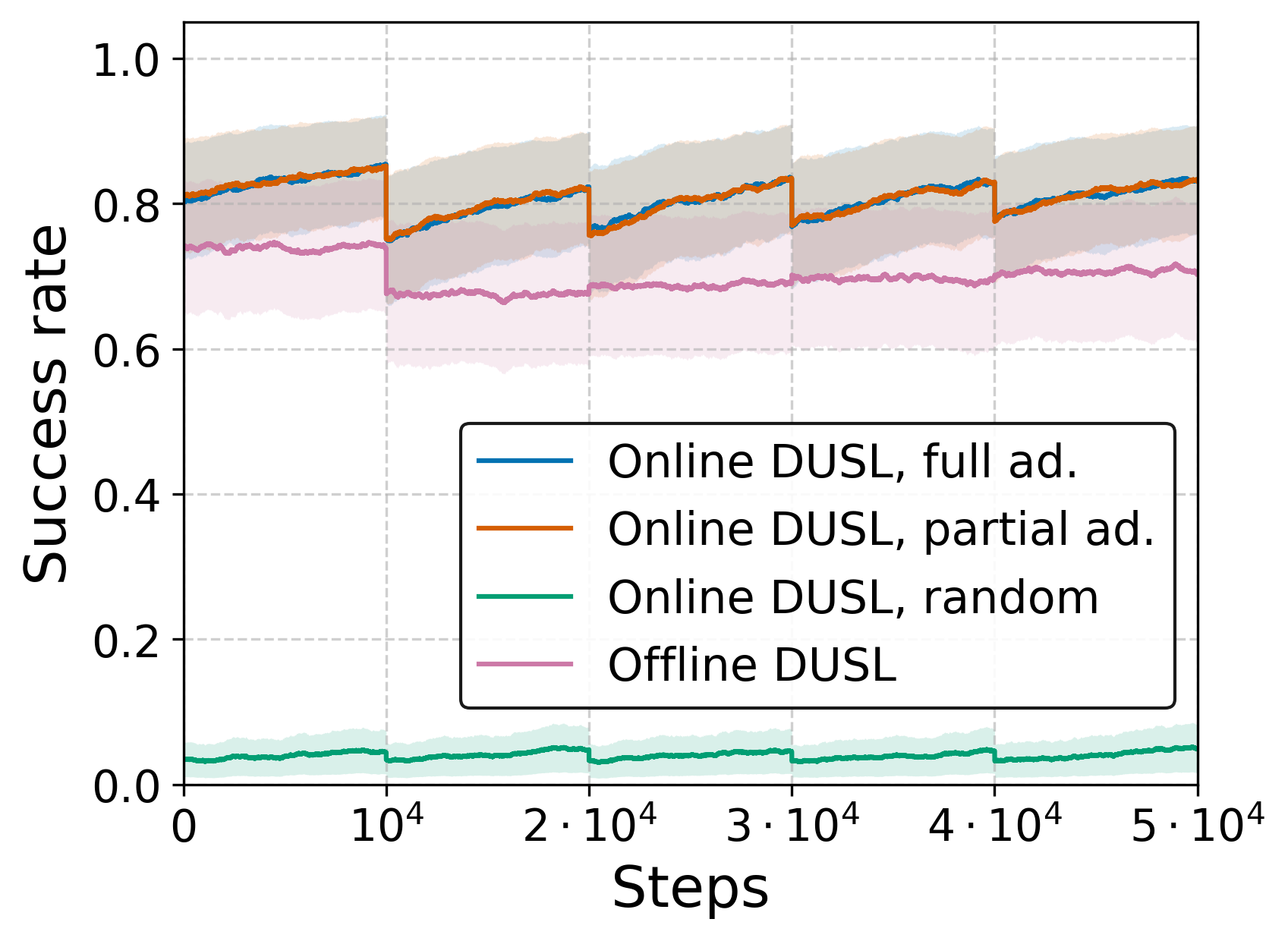}
        \caption{$N = 64$, $M = 64$, \\ $|\mathcal{A}_l| = 2$, $L = 4$.}
        \label{fig:general_fft_4_msg_fft}
    \end{subfigure}
    \caption{Adaptation to dynamic scenarios with general activation patterns and multiple messages, where 5 changes of $p_A$ occur.}
    \label{fig:general_fft_2-4_msg}
\end{figure}

\textit{\textbf{Dynamic scenarios.}} Finally, Fig.~\ref{fig:general_fft_2-4_msg} shows the performance of DUSL in dynamic environments 
for general activation patterns using the proposed online learning mechanism.
We 
set $N$ and $M$ as in Fig.~\ref{fig:general_off_2-4_msg}, and evaluate whether the online learning mechanism 
adapts quickly to shifts in the probability distribution \( p \). 
In these tests, the online learning mechanism proves to be effective to transfer knowledge. 
The online DUSL initialized from offline trained models, either with full adaptation or partial adaptation, exhibits the best adaption performance.  
The online DUSL initialized 
from scratch requires significantly more time to converge, whereas the offline DUSL without online learning mechanism performs worst with much lower success rates. 
In addition, we note from 
these results 
that fine-tuning through partial adaptation is generally the most computationally efficient solution, as it requires less operations to be performed online due to partial layer freezing in \acp{DNN} and achieves comparable performance to full adaptation. 

\blue{It is worth noting that in Figs. \ref{fig:general_off_2-4_msg} and \ref{fig:general_fft_2-4_msg}, the success rate does not reach the value one. This is mainly driven by the inherent difficulty of the general activation scenarios. Although longer training phase narrows the gap to the maximum attainable success rate, the ceiling remains below 1. This is because unstructured general activation patterns can have realizations $\mathcal{A}$ for which no collision-free assignment exists. 
We verified that extending the training time significantly narrows the performance gap. For the $L = 2$ scenario, extending the training from $6 \cdot 10^4$ to $4.8 \cdot 10^5$ steps improves the average success rate from 0.93 to 0.98, plateauing after $2.4 \cdot 10^5$ steps. Similarly, for the $L = 4$ scenario, extending training from $5 \cdot 10^5$ to $2 \cdot 10^6$ steps raises the average success rate from 0.89 to 0.94. This confirms that while performance improves with longer training, convergence to a perfect success rate is ultimately precluded by the scenario's inherent difficulty.}

\blue{Because active sets are generated probabilistically without structural dependencies, certain realized combinations of active sets may fundamentally lack a collision-free assignment under strict resource limits. For example, extending the scenario in Fig. \ref{fig:repetition_example} to include a possible active set $\{n_1, n_2, n_3\}$ would cause the previously successful repetition strategy to fail. Specifically, collisions would occur on both opportunity 1 and 2, preventing the delivery of messages $l_1$ and $l_2$. Encountering such unsolvable active sets caps the theoretical maximum success rate below one. 
}

\subsection{\blue{Online Learning Computational Complexity}}
\label{subsec:comp_complexity}

\blue{
Although DUSL integrates an online learning mechanism, it is specifically designed to adhere to strict latency and energy budgets of URLLC and constrained IoT deployments. This is achieved through the architectural decoupling of the \emph{real-time transmission path} (inference) and the \emph{asynchronous online tuning path} (gradient update).} \blue{
In URLLC, transmission latency is among the most critical constraints. When a node needs to transmit during the online learning phase, it performs a single forward pass of the underlying DNN architecture to determine its transmission opportunities. The DNN architecture is intentionally shallow, comprising an input layer, two hidden layers with 256 and 128 neurons, and an output layer of $M$ neurons. For a scenario with $M=4$, a single forward pass requires roughly 41.4k floating-point operations (FLOPs). Scaling the opportunities to $M=64$ marginally increases this to approximately 49.6k FLOPs. Experimental profiling of this forward pass on a standard edge device (Raspberry Pi 4) yields execution times consistently under 1 ms. Thus, the latency introduced by DUSL during the active transmission phase is practically negligible.} \blue{The online learning update relies on gradient descent, which requires a computationally heavier backward pass. Crucially, this update is triggered \emph{asynchronously} only after the central controller's ACK feedback is received; it runs in the background and does not block subsequent transmissions. Furthermore, to accommodate the strict energy budgets of IoT nodes, DUSL leverages \emph{partial adaptation}. By freezing the initial hidden layers and only updating the final layer's weights, the backward pass complexity is drastically reduced. Given that modern low-power microprocessors operate at an energy cost of picojoules per FLOP, the local energy consumed by the forward pass and partial backward pass translates to the range of microjoules. This local computational cost is orders of magnitude lower than the energy required to execute retransmissions (typically measured in millijoules) caused by a collision, ensuring DUSL remains highly energy-efficient in dynamic environments.
}

\section{Conclusion}
\label{sec:conclusions}

In this work, we introduced a novel decentralized \ac{MAC} protocol learning framework considering the transmission of multiple shared messages under limited transmission opportunities and dynamic node activation patterns. We proposed DUSL, a deep learning-based, unsupervised framework, to address the inherent coordination challenge due to the lack of inter-node communication. We provided theoretical insights into the problem of decentralized coordination, proving the optimality of deterministic strategies. We also derived bounds on the degradation of success probability due to changes in node activation probabilities, which is crucial for understanding system robustness in dynamic environments. Through extensive simulations, we demonstrated the scalability and adaptability of our proposed framework, and its superiority to baselines like distributed \ac{MAB}. These results validate our approach as an efficient solution for learning low-complexity, scalable, and reliable \ac{MAC} protocols, particularly for \ac{URLLC} scenarios and energy-constrained \ac{IoT} applications.

\blue{While the current DUSL framework demonstrates strong adaptability to dynamic activation distributions within a fixed network size, scaling directly to networks with a varying total number of nodes $N$ remains an open question that we leave for future research. 
Currently, DUSL operates as a heterogeneous model, meaning each node is equipped with its own specialized neural networks. To achieve zero-shot or few-shot scalability to different network sizes, we would need to transition to a homogeneous model architecture. 
{In such as design, all nodes can deploy a shared base model to synthesize transmission strategies, and each node only needs to fine-tune a local read-out layer with our online learning mechanism to adapt to networks with different sizes. We leave the design and evaluation of this homogeneous architecture for future work.}
}

\blue{Furthermore, to completely eliminate the computational and energy overhead of on-device online learning, a promising avenue for future work is the integration of a network digital twin (DT)-driven closed-loop update mechanism. While our current approach models general correlation across transmitting nodes using a Dirichlet distribution, a DT would allow for incorporating a prior or an explicit model of the underlying physical phenomenon observed by the network. In this paradigm, the central server would utilize lightweight responses from the \ac{IoT} nodes to continuously calibrate the DT. 
All training operations (i.e., stochastic exploration, reward computation, and policy updates) would be offloaded entirely to this DT. Following convergence, the server would periodically deploy the  deterministic policies back to the \ac{IoT} nodes, restricting the devices' operations only to deterministic inference.}

\appendices

\section{Proof of Theorems \ref{th:1}, \ref{thm2}, and \ref{th:3}}
\label{appendix:A}

% Proof 1
\noindent \textit{Proof of Theorem \ref{th:1}.}
The optimization problem in (\ref{eq:optim}) is standard linear program
\begin{align}
\begin{split}
\label{eq:stand_lp}
    \maxb_{\mathbf{\Psi}}~\mathbb{E}[\,\xi \mid \mathbf{\Psi}\,] = \maxb_{\bm{\psi}}~\bm{\omega}^T \bm{\psi}
    ~,
\end{split}
\end{align}
subject to simplex constraints $\bm{\psi} \geq \mathbf{0}$ and $\mathbf{1}^T\bm{\psi} = 1$, since
\begin{align}
\begin{split}
    \mathbb{E}[\,\xi \mid \mathbf{\Psi}\,] &= \!\!\!\!\!\!\!\!\!\!\sum_{\mathbf{X} \in \{0, 1\}^{L \times N \times M}}\!\!\!\!\!\!\!\!\!\!\mathbb{E}[\,\xi \mid \mathbf{\Psi} = \mathbf{X}\,] \\
    &= \!\!\!\!\!\!\!\!\!\!\sum_{\mathbf{X} \in \{0, 1\}^{L \times N \times M}}\!\!\!\!\!\!\!\!\!\! \mathbf{\psi}(\mathbf{X}) \sum_{\mathcal{A} \in \mathcal{R}_{A}} p_{A}(\mathcal{A}) ~\xi(\mathcal{A}, \mathbf{X}) \\
    &=\!\!\!\!\!\!\!\!\!\!\sum_{\mathbf{X} \in \{0, 1\}^{L \times N \times M}}\!\!\!\!\!\!\!\!\!\! \mathbf{\psi}(\mathbf{X})~\omega(\mathbf{X}) = \bm{\omega}^T \bm{\psi}
    ~,
\end{split}
\end{align}
where $\omega(\mathbf{X}) = \sum_{\mathcal{A} \in \mathcal{R}_{A}} p_{A}(\mathcal{A}) ~\xi(\mathcal{A}, \mathbf{X}) \in [0, 1]$ is the probability of successful transmission given $\mathbf{X}$. The Lagrangian of (\ref{eq:stand_lp}) is
\begin{align}
\begin{split}
    L(\bm{\psi}, \bm{\lambda}, \nu) &= \bm{\omega}^T \bm{\psi} - \bm{\lambda}^T \bm{\psi} + \nu(\mathbf{1}^T\bm{\psi} - 1) \\
    &= -\nu + (\bm{\omega} - \bm{\lambda}+\nu \mathbf{1})^T \bm{\psi}
    ~,
\end{split}
\end{align}
where $\bm{\lambda}$ and $\nu$ are the Lagrangian multipliers, the former for the $2^{KM}$ inequality constraints, the latter for the equality constraint. Maximizing with respect to $\bm{\psi}$ gives the dual function 
\begin{align}
\begin{split}
    g(\bm{\lambda}, \nu) = \sup_{\bm{\psi}} L(\bm{\psi}, \bm{\lambda}, \nu) = -\nu
    ~,
\end{split}
\end{align}
where $\sup_\mathcal{\bm{\psi}}$ is the supremum of $\bm{\psi}$ and $\bm{\omega} - \bm{\lambda} + \nu \mathbf{1}= \mathbf{0}$ is the dual feasibility condition, otherwise $g(\bm{\lambda}, \nu)$ is not bounded from above as $L$ is a linear function of $\bm{\psi}$. By considering the dual feasibility condition, the dual problem is therefore
\begin{align}
\begin{split}
    \minb_{\bm{\lambda}, \nu} g(\bm{\lambda}, \nu) = -\nu~, \text{ s.t. } \bm{\lambda} = \bm{\omega} + \nu \mathbf{1} \leq \mathbf{0}
    ~.
\end{split}
\end{align}
Here, from the dual constraint we get $- \nu \mathbf{1} \geq \bm{\omega}$, that is a vector inequality, equivalent to $\forall\,\mathbf{X} \in \{0, 1\}^{L \times N \times M}, - \nu \geq \omega(\mathbf{X})$, and consequently $- \nu \geq \maxb \omega(\mathbf{X})$, so the dual solution is found at $\maxb \omega(\mathbf{X})$. We note that since  the primal problem (\ref{eq:stand_lp}) is a linear optimization problem, strong duality holds, therefore the primal and the dual optimal objectives are equal:
\begin{align}
\begin{split}
    \maxb_{\mathbf{\Psi}}\,\mathbb{E}[\,\xi \mid \mathbf{\Psi}\,] = \maxb_{\bm{\psi}}\,\bm{\omega}^T \bm{\psi} = \minb_{\bm{\lambda}, \nu} g(\bm{\lambda}, \nu) = \maxb_{\mathbf{X}}\omega(\mathbf{X})
    ~.
\end{split}
\end{align}
If $\mathbf{X}^* = \argmax\,\omega(\mathbf{X})$ is unique, then 
\begin{align}
\begin{split}
    \maxb_{\mathbf{\Psi}}~\mathbb{E}[\,\xi \mid \mathbf{\Psi}\,] = \omega(\mathbf{X}^*)~\text{ and }~\mathbf{\psi}(\mathbf{X}^*) = 1
    ~,
\end{split}
\end{align}
i.e., $\mathbf{\Psi}$ is the only optimal joint strategy and it is deterministic. On the other hand, if there is a set $\{\mathbf{X}_i^* \}^D_{i=1}$ such that $\mathbf{X}^*_i = \argmax\,\omega(\mathbf{X}) ~\forall i \in \{1, \ldots, D\}$, then
\begin{align}
\begin{split}
   \maxb_{\mathbf{\Psi}}~\mathbb{E}[\,\xi \mid \mathbf{\Psi}\,] = \omega(\mathbf{X}^*_i)~\text{ and }~\sum_{i = 1}^D \mathbf{\psi}(\mathbf{X}^*_i) = 1
    ~,
\end{split}
\end{align}
i.e., there exists a finite set $\{\mathbf{\Psi}_i' \}^D_{i=1}$ of optimal deterministic joint strategies, where $\mathbf{\psi}_i'(\mathbf{X}^*_i) = 1 \text{ and } \mathbf{\psi}_i'(\mathbf{X}^*_k) = 0~\forall i, k \in \{1, \ldots, D\}, i \neq k$, and an infinite set $\{\mathbf{\Psi}_j'' \}^\infty_{j=1}$ of optimal randomized joint strategies, where $\mathbf{\psi}_j''(\mathbf{X}^*_k) < 1~\forall j, k \in \{1, \ldots, D\}$.
\begin{flushright}$\square$\end{flushright}

% Proof 2
\deniz{
\noindent \textit{Proof of Theorem \ref{thm2}.}
We prove the NP-hardness of the optimization problem in (\ref{eq:optim}) by a polynomial-time reduction from the Maximum Cut (MAX-CUT) problem, which is known to be NP-hard \cite{karp2009reducibility}. 
To prove that an optimization problem is NP-hard, it is sufficient to show that a restricted special case of the problem is NP-hard. While our system model allows for $M \geq L$ communication opportunities and arbitrary active sets $\mathcal{A}_l$, we restrict our construction to the specific instance where the number of opportunities exactly equals the number of messages ($M = L$), and where the active subsets emulate graph edges ($|\mathcal{A}_l| = 2$). 
Consider an arbitrary instance of the MAX-CUT problem defined by an undirected graph $G=(V, E)$. The objective is to partition the vertices $V$ into two disjoint sets to maximize the number of crossing edges. We map this to our transmission problem as follows:
\begin{itemize}
    \item Set the number of nodes $N = |V|$, where each node $n_i \in \mathcal{N}$ uniquely corresponds to a vertex $v_i \in V$.
    \item Define the probability mass function $p(\mathcal{A})$ to be uniform over $|E|$ specific patterns. Each pattern $\mathcal{P}_e$ corresponds to an edge $e = (u,v) \in E$, such that $p(\mathcal{P}_e) = \frac{1}{|E|}$.
    \item For a given pattern $\mathcal{P}_e$ corresponding to edge $e = (u,v)$, we set nodes $u$ and $v$ to be active for all $L$ messages. That is, $\mathcal{A}_l = \{u, v\}$ for every $l \in \{1, \ldots, L\}$.
\end{itemize}
By Theorem \ref{th:1}, we restrict our search to deterministic joint strategies $\mathbf{X}$ without loss of optimality. For a pattern $\mathcal{P}_e$ to be successful, meaning the indicator $\xi(\mathcal{P}_e, \mathbf{X}) = 1$, two conditions must hold for \textit{every} message $l$: exactly one active node must transmit message $l$ on some opportunity $m_l$ ($\xi_a(l,m_l) = 1$), and no other message can be transmitted on $m_l$ ($\xi_b(l,m_l) = 1$). 
Because our restricted instance enforces $M = L$, condition $\xi_b$ requires the $L$ messages to be bijectively mapped to the $M$ opportunities to avoid inter-message collisions. Without loss of generality, we can fix this mapping such that message $l$ is exclusively assigned to opportunity $l$. 
Under this mapping, the local strategy of node $n$ simplifies to a binary decision: $y_{l,n} \in \{0,1\}$, where $y_{l,n} = 1$ indicates node $n$ transmits message $l$, and $y_{l,n} = 0$ indicates it remains silent. To satisfy $\xi_a(l,l) = 1$ for nodes $u$ and $v$, exactly one node must transmit while the other remains silent (i.e., $y_{l,u} \neq y_{l,v}$). 
Thus, the optimization objective evaluates to:
\begin{equation}
    \max_{\mathbf{X}} \mathbb{E}[\xi \mid \mathbf{\Psi}, p] = \max_{\mathbf{y}} \frac{1}{|E|} \sum_{(u,v) \in E} \left( \bigwedge_{l=1}^{L} I(y_{l,u} \neq y_{l,v}) \right)~.
\end{equation}
The logical conjunction dictates that an edge pattern is only successful if nodes $u$ and $v$ make opposite decisions for every message. 
To maximize this objective, the optimal deterministic strategy must maintain consistent transmission assignments across all $L$ messages (i.e., setting $y_{1,n} = y_{2,n} = \dots = y_{L,n} = y_n$ for each node $n$). Consequently, the objective reduces to maximizing $\sum_{(u,v) \in E} I(y_u \neq y_v)$.
Assigning $y_n = 1$ (transmitting) places vertex $v_n$ in one MAX-CUT partition, and $y_n = 0$ (silent) places it in the other. Finding the optimal joint transmission strategy exactly yields the maximum cut of the graph $G$. Because this reduction maps any MAX-CUT instance to our problem in polynomial time $O(|V| + L|E|)$, the restricted instance is NP-hard. By the principle of restriction, the generalized optimization problem in (\ref{eq:optim}) for $M \geq L$ and arbitrary active sets $\mathcal{A}_l$ is therefore NP-hard.
\begin{flushright}$\square$\end{flushright}
}

% Proof 3
\noindent \textit{Proof of Theorem \ref{th:3}.}
Defining by $\mathbf{X}$ the moves of all the nodes according to $\mathbf{\Psi}$, we can represent the set of values of $A$ for which the condition for a successful transmission holds as  
\begin{align}
\begin{split}
    \mathcal{S} = \left\{ \mathcal{A}~:~ \xi(\mathcal{A}, \mathbf{X}) = 1,~\forall \mathcal{A} \in \mathcal{R}_{A}\right\}~,
\end{split}
\end{align}
and consequently we get the expected probability of successful transmission as $\xi' =  \sum_{\mathcal{A} \in \mathcal{S}} p_{A}'(\mathcal{A})$ and $\xi'' =  \sum_{\mathcal{A} \in \mathcal{S}} p_{A}''(\mathcal{A})$. Using the aggregation property of the Dirichlet distribution, we can model $(\xi', 1 - \xi')$ and $(\xi'', 1 - \xi'')$ as samples of a Dirichlet distribution of order two: 
\begin{align}
\begin{split}
    \left(\xi', 1 - \xi'\right),\,\left(\xi'', 1 - \xi''\right) &\sim \text{Dir}\left(\sum_{\mathcal{A} \in \mathcal{S}} \alpha_\mathcal{A}\,, \sum_{\mathcal{A} \in \mathcal{R}_{A} \setminus \mathcal{S}} \alpha_\mathcal{A}\right) \\
    &= \text{Dir}\left(\alpha_S\,, \alpha_{F} \right)~,
\end{split}
\end{align}
where $\alpha_\mathcal{A}$ is the concentration parameter associated to both $p_{A}'(\mathcal{A})$ and $p_{A}''(\mathcal{A})$, $\forall \mathcal{A} \in \mathcal{R}_{A}$, and $\alpha_S = \sum_{\mathcal{A} \in \mathcal{S}} \alpha_\mathcal{A}$ and $\alpha_F = \sum_{\mathcal{A} \in \mathcal{R}_{A} \setminus \mathcal{S}} \alpha_\mathcal{A}$ are concentration parameters of a Dirichlet distribution of order two. Since the marginal distributions of a Dirichlet distribution are Beta distributions, we know that $\xi'$ and $\xi''$ follow a Beta distribution; that is, $\xi', \xi'' \sim \text{Beta}\left(\alpha_S\,, \alpha_{F} \right)$. 
\deniz{The variance of the difference $\mathrm{Var}(\xi' - \xi'')$ can be written as
\begin{equation}
\begin{aligned}
    \mathrm{Var}(\xi' - \xi'') &= \mathrm{Var}(\xi') + \mathrm{Var}(\xi'') - 2\,\mathrm{Cov}(\xi',\,\xi'') \\
    &= \frac{2\,\alpha_S\,\alpha_F}{(\alpha_S + \alpha_F)^2\,(1 + \alpha_S + \alpha_F)}~,
\end{aligned}
\end{equation}
where the covariance is zero given the i.i.d. assumption on $p_A'$ and $p_A''$. By setting $\alpha = \alpha_S + \alpha_F$, we can bound $\mathrm{Var}(\xi' - \xi'')$ using the inequality of arithmetic and geometric means. Since $\frac{\alpha_S + \alpha_F}{2} \geq \sqrt{\alpha_S \alpha_F}$, it follows that $2 \alpha_S \alpha_F \leq \frac{\alpha^2}{2}$. Therefore:
\begin{equation}
\begin{aligned}
    \mathrm{Var}(\xi' - \xi'') &= \frac{2\,\alpha_S\,\alpha_F}{\alpha^2\,(1 + \alpha)} \\
    &\leq \frac{\alpha^2 / 2}{\alpha^2\,(1 + \alpha)} = \frac{1}{2\,(1 + \alpha)} \leq \frac{1}{2\,(1 + \alpha_{\varepsilon}\,2^{K})}~.
\end{aligned}
\end{equation}
}The final inequality holds as $\alpha = \sum_{i=1}^{2^K} \alpha_i \geq \alpha_{\varepsilon} \cdot 2^{K}$, where $\alpha_{\varepsilon} = \min \{\alpha_i \}^{2^K}_{i=1}$. 
We can then obtain (\ref{eq:th3}) by applying Chebyshev's inequality:
\deniz{
\begin{equation}
\begin{aligned}
    P(\vert \xi' - \xi'' \vert \geq \eta) &\leq \frac{\mathrm{Var}(\xi' - \xi'')}{\eta^2} \\
    &\leq \frac{1}{2\,\eta^2\,(1 + \alpha_{\varepsilon}\,2^{K})}~.
\end{aligned}
\end{equation}
}Finally, since $\xi'$ and $\xi''$, and consequently $\vert \xi' - \xi'' \vert$, are bounded in $[0, 1]$, it suffices to consider $\eta \in (0, 1]$.
\begin{flushright}$\square$\end{flushright}

\section{Additional Experiments}
\label{appendix:B}

\subsection{Unbalanced Message Availability}
\label{sec:unbalanced}

First, we investigate the performance of DUSL in a scenario with unbalanced message availability. We adopt the conditional activation pattern setting with $N = 64$ nodes, $M = 32$ communication opportunities, and $L=2$ messages. The imbalance is introduced in the cardinality of the active sets: at each time step, the first message ($l_1$) is available to a set of $|\mathcal{A}_1| = 8$ nodes, while the second message ($l_2$) is available to only a single node, i.e., $|\mathcal{A}_2| = 1$. 
The results are presented in Fig. \ref{fig:unbalanced}. As shown, the DUSL framework successfully learns an effective transmission protocol, with all variants of DUSL significantly outperforming the random baseline. In particular, the high-exploration case ($\varepsilon=0.0$) converges the fastest, while the low-exploration case ($\varepsilon=1.0$) learns more slowly but still reaches a comparable performance level, indicating that DUSL is robust to asymmetries in message distribution. 

\subsection{Temporal Correlation in Activation Patterns}
\label{sec:temporal}

\begin{figure}[t]
    \centering
    \begin{subfigure}[b]{0.49\columnwidth}
        \centering
        \captionsetup{justification=centering}
        \includegraphics[width=\textwidth]{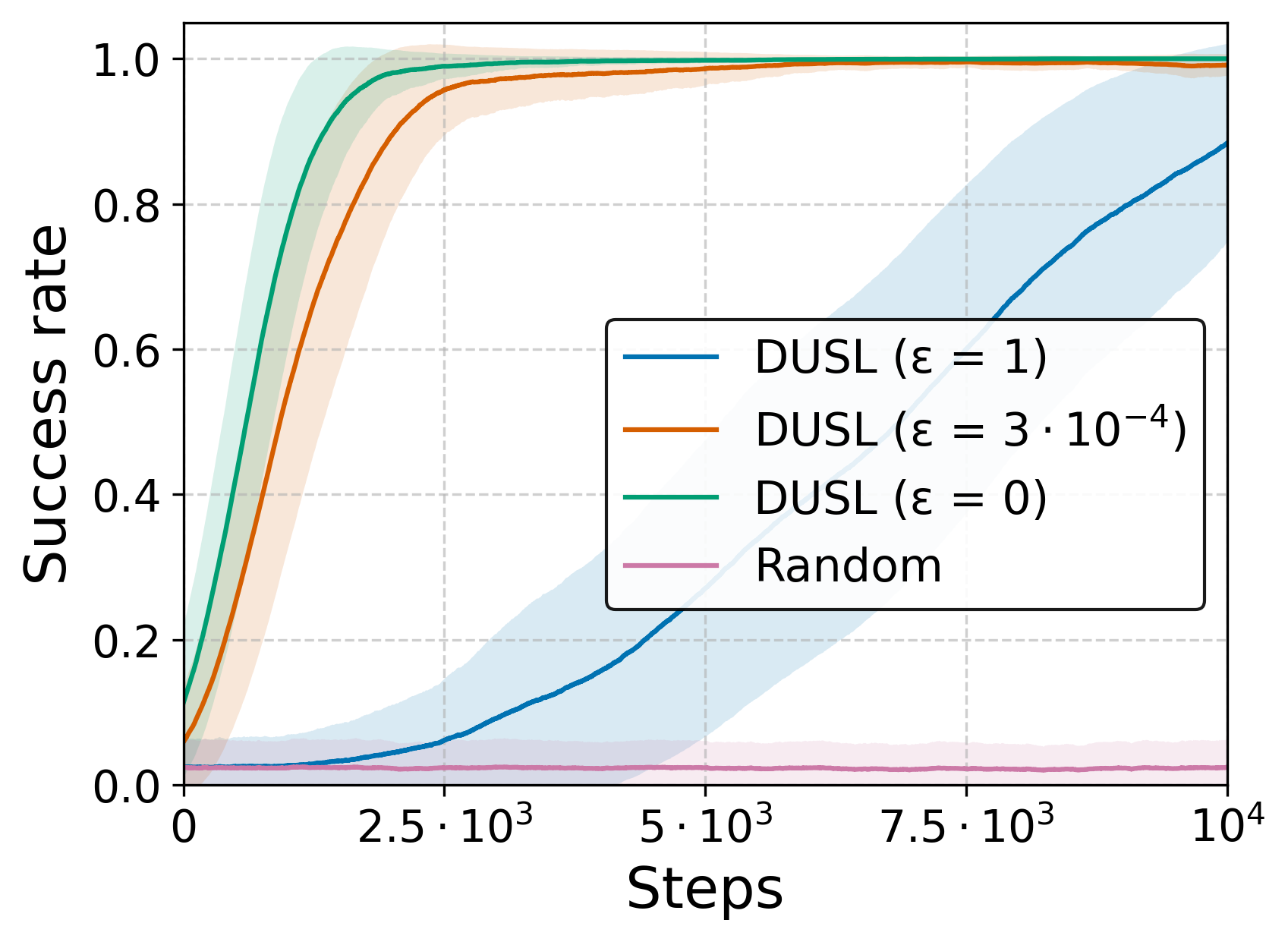}
        \caption{$N = 64$, $M = 32$, \\ $|\mathcal{A}_1| = 8$, $|\mathcal{A}_2| = 1$.}
        \label{fig:unbalanced}
    \end{subfigure}
    \hfill
    \begin{subfigure}[b]{0.49\columnwidth}
    \captionsetup{justification=centering}
        \includegraphics[width=\textwidth]{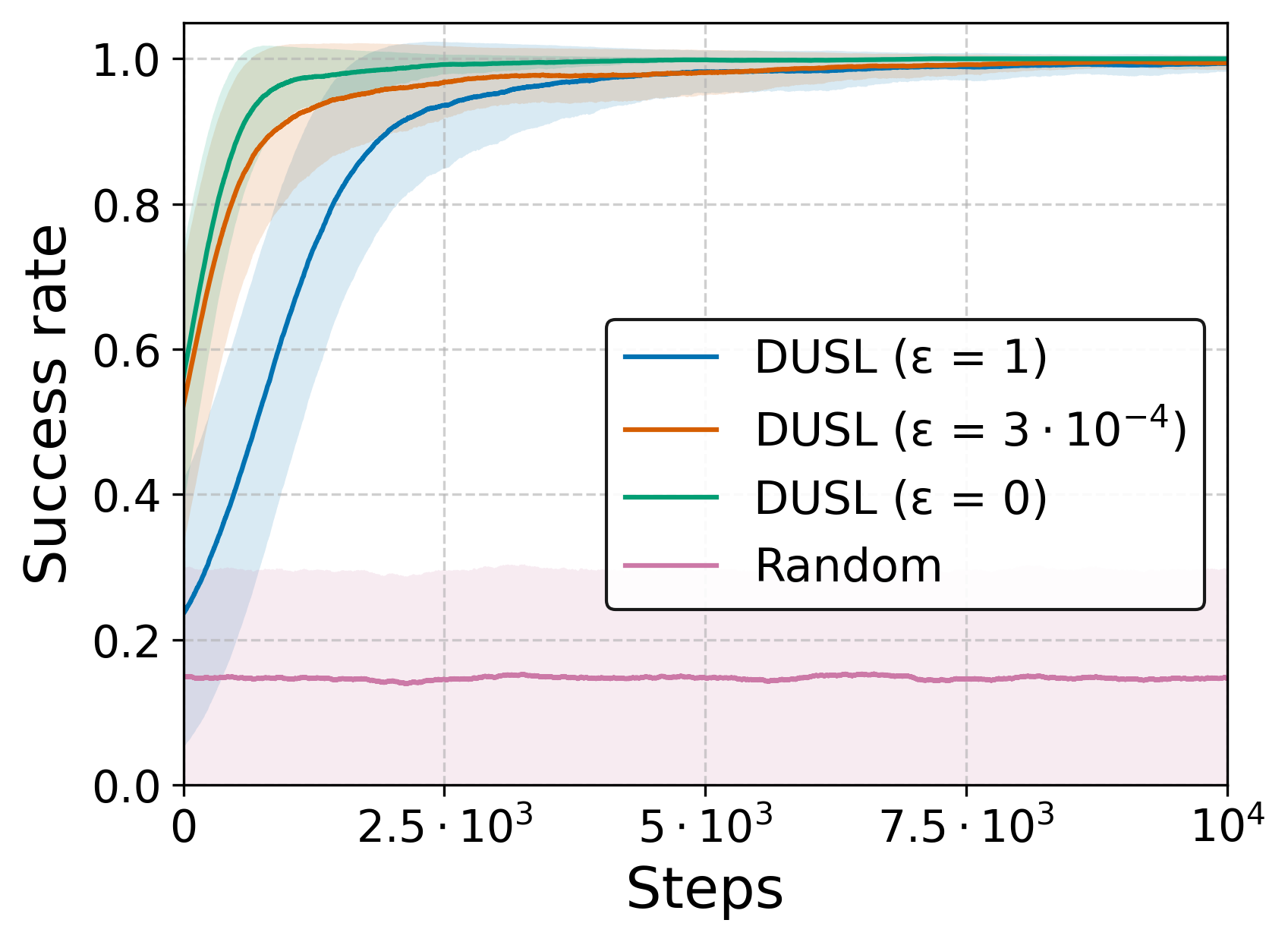}
        \caption{$N = 64$, $M = 32$, \\ $|\mathcal{A}_l| = 4$, $L = 2$, $b_{\max}=10$.}
        \label{fig:temporal}
    \end{subfigure}
    \caption{(a) Training in a static scenario with conditional activation patterns and unbalanced message availability, and (b) training in a static scenario with temporally correlated conditional activation patterns.}
    \label{fig:appendix_b}
\end{figure}

In the second experiment, we study the impact of temporal correlation in the activation patterns. We extend the conditional activation scenario by introducing a mechanism where the active sets can persist over multiple time steps. Specifically, for each message $l$ at each time step $t$, after a new active set $\mathcal{A}_l(t)$ is sampled, we also sample an integer persistence duration $b_{t,l}$ uniformly from $\{0, \ldots, b_{\max}\}$. The active set $\mathcal{A}_l(t)$ then remains unchanged for the next $b_{t,l}$ time steps, i.e., $\mathcal{A}_l(t) = \mathcal{A}_l(t+1) = \dots = \mathcal{A}_l(t+b_{t,l})$. For this experiment, we set $N = 64$, $M = 32$, $L=2$, $|\mathcal{A}_l| = 4$ for both messages, and $b_{\max} = 10$. 
The results, shown in Fig. \ref{fig:temporal}, demonstrate that DUSL is 
robust to such 
temporal correlation as well. DUSL quickly adapts and converges to a high success rate, whereas the random policy's performance remains low. 
This suggests that the DUSL framework effectively operate in environments where the underlying state 
exhibits temporal dependencies.

\subsection{Learned Policy Statistics}
\label{sec:policy_stats}

\begin{figure}[t]
    \centering
    
    \begin{subfigure}[b]{0.95\columnwidth}
        \centering
        \includegraphics[width=\columnwidth]{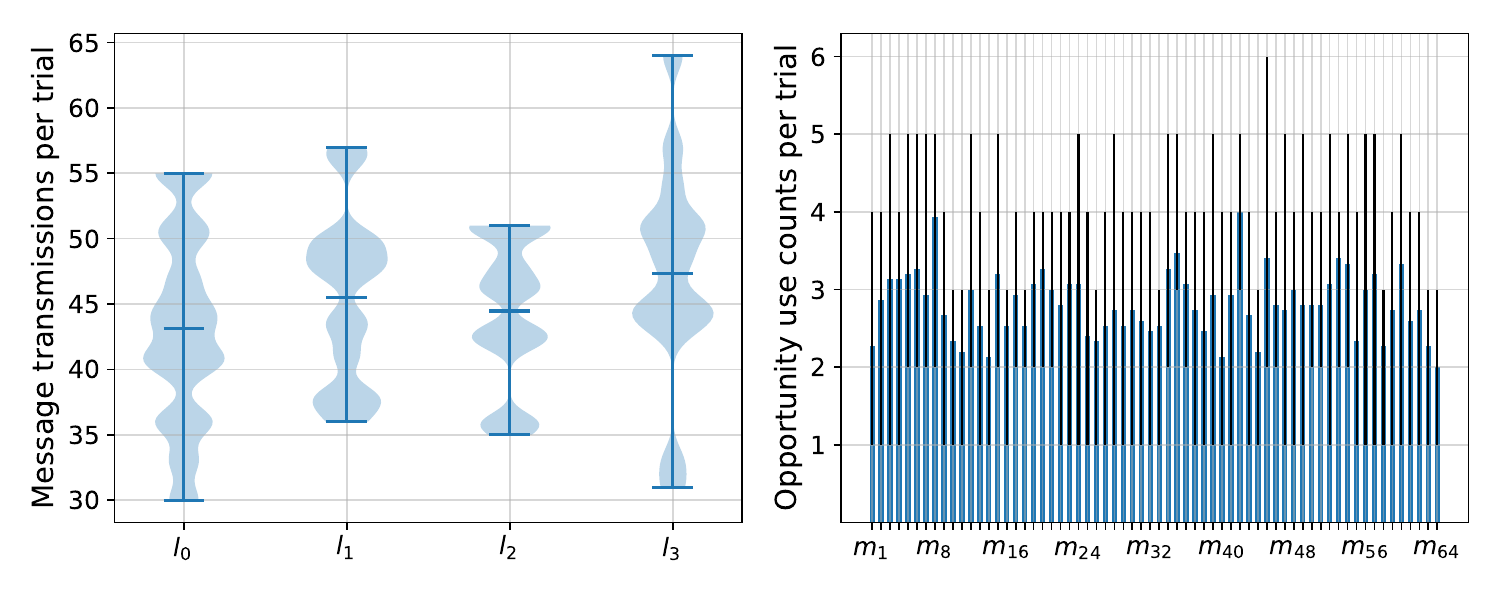}
        \caption{Message transmissions and opportunity use counts per trial in conditional activation scenario.}
        \label{fig:stats_c0064}
    \end{subfigure}
    \vfill
    \begin{subfigure}[b]{0.95\columnwidth}
        \centering
        \includegraphics[width=\columnwidth]{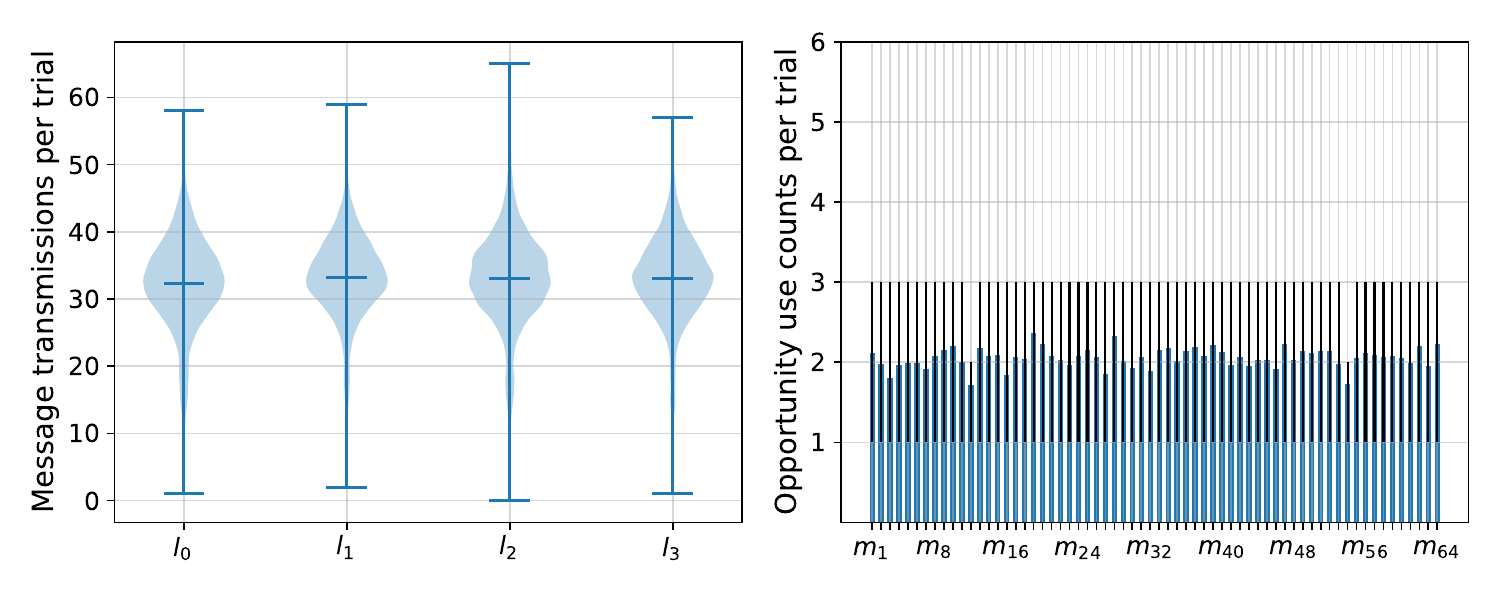}
        \caption{Message transmissions and opportunity use counts per trial in general activation scenario.}
        \label{fig:stats_s0064}
    \end{subfigure}
    
    \caption{Statistical analysis of learned policies for $N=64, M=64, L=4, |\mathcal{A}_l|=2$.}
    \label{fig:policy_stats}
\end{figure}

To better understand the strategies developed by DUSL, we now investigate the statistics of the learned policies for two representative scenarios: the conditional activation case from Fig.~\ref{fig:cond_off_4_msg} and the general activation case from Fig.~\ref{fig:general_off_4_msg}. Both scenarios use $N=64, M=64, L=4,$ and $|\mathcal{A}_l|=2$.
Figure~\ref{fig:policy_stats} presents these statistics. The left plots of Fig.~\ref{fig:stats_c0064} and Fig.~\ref{fig:stats_s0064} show the distribution of the total number of message transmissions per trial for each of the $L=4$ messages, presented as violin plots. In both scenarios, the policies %clearly 
leverage message repetitions to ensure delivery. This can be observed %is evident 
from the wide distributions and that the expected transmission counts significantly greater than the number of active nodes ($|\mathcal{A}_l|=2$). 
This demonstrates that the algorithm learns to build significant robustness into its strategy to maximize the success probability.

The right plots in Fig.~\ref{fig:policy_stats} show the use counts of communication opportunities, displaying the mean usage (blue bar) and the 25-75 percentile range (black bars) for 
the $M=64$ opportunities. A key finding is that for both learned policies, the 25th percentile for opportunity usage is in most cases 1. This is a critical outcome: it signifies that in at least 25\% of the trials, these opportunities experience exactly one transmission. As a use count of 1 represents a successful, collision-free transmission, the fact that the learned policies consistently ensure this condition is met across a wide range of opportunities is a primary reason for their high success rates.
Importantly, successful decoding does not require every opportunity to be collision-free, but only requires  
the $L$ messages being delivered at least once. In other words, it suffices that there exist at least $L$ out of the $M$ opportunities that are used correctly (i.e., produce a collision-free transmission of the corresponding message), while the remaining $M-L$ slots may contain %redundant repeats or 
collisions without harming overall success, %. 
which explains why the policies can tolerate a certain number of collisions yet still achieve high success rates.

\begingroup
\small
\bibliographystyle{IEEEtran}
\bibliography{references}
\endgroup

\end{document}